\documentclass[amsmath,amssymb,aps,pre,showkeys,reprint,eqsecnum]{revtex4-1}

\usepackage{graphicx}
\usepackage{dcolumn}
\usepackage{bm}

\begin{document}

\title{Logarithmic violation of scaling in strongly anisotropic
turbulent transfer of a passive vector field}

\author{N. V. Antonov and N. M. Gulitskiy}

\email{n.antonov@spbu.ru, ngulitskiy@gmail.com}

\affiliation{Chair of High Energy Physics and Elementary Particles \\
Department of Theoretical Physics, Faculty of Physics \\
Saint Petersburg State University, Ulyanovskaja~1 \\
Saint~Petersburg--Petrodvorez, 198504 Russia}


\begin{abstract}
Inertial-range asymptotic behavior of a vector (e.g., magnetic) field,
passively advected by a strongly anisotropic turbulent flow, is studied
by means of the field theoretic renormalization group and the operator
product expansion. The advecting velocity field is Gaussian, not correlated
in time, with the pair correlation function of the form
$\propto \delta(t-t') / k_{\bot}^{d-1+\xi}$, where
$k_{\bot}=|{\bf k}_{\bot}|$ and  ${\bf k}_{\bot}$
is the component of the wave vector, perpendicular to the distinguished
direction (``direction of the flow'') -- the $d$-dimensional generalization
of the ensemble introduced by Avellaneda and Majda
[{\it Commun. Math. Phys.} {\bf 131}: 381 (1990)].
The stochastic advection-diffusion equation for the transverse
(divergence-free) vector field includes, as special cases, the kinematic
dynamo model for
magnetohydrodynamic turbulence and the linearized Navier--Stokes equation.
In contrast to the well known isotropic Kraichnan's model, where various
correlation functions exhibit anomalous scaling behavior with infinite sets
of anomalous exponents, here the dependence
on the integral turbulence scale $L$ has a logarithmic behavior: instead
of power-like corrections to ordinary scaling, determined by naive
(canonical) dimensions, the anomalies manifest themselves as polynomials
of logarithms of $L$. The key point is that
the matrices of scaling dimensions of the relevant families
of composite operators appear nilpotent and cannot be diagonalized.
The detailed proof of this fact is given for the correlation functions of
arbitrary order.

\end{abstract}



\maketitle

\section{Introduction} \label{sec:Intro}

Much attention has been attracted to the problem of intermittency and
anomalous scaling in developed magnetohydrodynamic (MHD) turbulence;
see, e.g.,~\cite{GM}--\cite{SW1} and references therein.
It has long been known that in the so-called Alfv{\'e}nic regime, the
MHD turbulence demonstrates the behavior, similar to that of the usual
fully developed fluid turbulence: cascade of energy from the infrared
range towards smaller scales, where the dissipation effects dominate,
and self-similar (scaling) behavior of the energy spectra in the
intermediate (inertial) range. Moreover, intermittent character of the
fluctuations in the MHD turbulence is much strongly pronounced than in
ordinary turbulent fluids.

The solar wind provides a kind of appropriate ``wind tunnel'' in which
different approaches
and models of the MHD turbulence can be tested~\cite{SW2}--\cite{SW6}. In
solar flares, highly energetic and anisotropic large-scale motions coexist
with small-scale coherent structures, finally responsible for the dissipation.
Thus modelling the way how the energy is redistributed, transferred along
the spectra and eventually dissipated is a difficult task. The intermittency
strongly modifies the scaling behavior of the higher-order correlation
functions, leading to anomalous scaling, described by infinite sets of
independent ``anomalous exponents.''

A simplified description of the situation was proposed in~\cite{E}: the
large-scale field $B^{0}_{i} = n_{i} B^{0}$ dominates the dynamics in the
distinguished direction ${\bf n}$, while the activity in the perpendicular
plane is described as nearly two-dimensional. This picture allows for
precise numerical simulations, which show that turbulent fluctuations
organize in rare coherent structures separated by narrow current
sheets. On the other hand, the observations and simulations show that the
scaling behavior in the solar wind is closer to the anomalous scaling in the
three-dimensional fully developed hydrodynamic turbulence, rather than to
simple Iroshnikov--Kraichnan scaling suggested by two-dimensional picture
with the inverse energy cascade; see, e.g., the discussion in~\cite{SW2}.
Thus further analysis of more realistic three-dimensional models is welcome.

Two main simplifications of the full-scale model are possible here. First,
the magnetic field can be taken {\it passive}, that is, not to affect
the dynamics of the velocity field. This approximation is valid when the
gradients of the magnetic fields are not too large. What is more, the
renormalization group
analysis shows that such a ``kinematic regime'' can indeed describe the
possible infrared (IR) behavior of the full-scale model~\cite{K}.

Second, description of the fluid turbulence remains itself a difficult task.
Once the feedback of the magnetic field is neglected, the velocity can be
modelled by statistical ensembles with prescribed properties.

In spite of their relative simplicity, the models of passive fields,
advected by such ``synthetic'' velocity ensembles, reproduce
many of the anomalous features of genuine turbulent mass or heat transport.
At the same time, they admit a detailed analytical treatment.
Most remarkable progress was achieved for Kraichnan's ``rapid-change model,''
where the correlation function of the velocity is taken in the power-like
form $\langle vv \rangle \propto \delta(t-t') k^{-d-\xi}$, where $k$ is the
wave number, $d$ is the dimension of space and $\xi$ is an arbitrary exponent.
There, for a passive {\it scalar} field (temperature or density of an
impurity), the existence of anomalous scaling was established on the basis
of a microscopic dynamic model~\cite{Kraich1}; the
corresponding exponents were calculated within controlled approximations
\cite{GK} and, eventually, within systematic perturbation expansions in a
formal small parameter $\xi$~\cite{RG}. Detailed review of the theoretical
research on the passive scalar problem and the bibliography can be found
in~\cite{FGV}.

Owing to the presence of a new stretching term (in addition to the advecting
term) in the dynamic equation for the vector (e.g., magnetic) field, the
behavior of the
passive {\it vector} fields appears much richer than for the scalar case
\cite{KA68}--\cite{HA}: they reveal anomalous scaling already on the level of
the pair correlation function~\cite{V96,RK97} and develop some large-scale
instabilities, interpreted as the turbulent dynamo effect
\cite{KA68,V96,DV}. Quoting~\cite{Legacy}: ``...there is considerably more
life in the large-scale transport of vector quantities'' (p. 232).

A most powerful method to study the anomalous scaling in various statistical
models
of turbulent advection is provided by the field theoretic renormalization
group (RG) and operator product expansion (OPE); see the monographs
\cite{Zinn,Vasiliev-Green} and references therein. In the RG+OPE
scenario~\cite{RG}, anomalous scaling emerges as a consequence of the
existence in the model of composite fields (``composite operators'' in
the quantum-field terminology) with {\it negative} scaling dimensions;
see~\cite{JphysA} for a review and the references. In a number of papers
\cite{Lanotte2,alpha,AntGul2012,Marian,AntGul2013} the RG+OPE approach
was applied to the case of passive vector (magnetic) fields in Kraichnan's
ensemble, and to its generalizations (large-scale anisotropy,
compressibility, finite correlation time, non-Gaussianity, more general form
of the nonlinearity). Explicit analytical expressions were derived for the
anomalous exponents to the first~\cite{Lanotte2,alpha} and the second
\cite{AntGul2012,Marian} orders in $\xi$.
For the pair correlation function of the magnetic field, exact results
were obtained within the zero-mode approach~\cite{V96,RK97,Lanotte}.

In this paper, we apply the RG+OPE approach to the inertial-range behavior
of strongly anisotropic
MHD turbulence within the framework of a simplified model, where the
magnetic field is passive and the velocity field is modelled by a Gaussian
ensemble with prescribed statistics. Our model differs from the conventional
Kazantsev--Kraichnan kinematic dynamo model in two respects:

(1) It involves a general relative coefficient ${\cal A}$ between the
stretching and the advecting terms in the equation for the vector field.
Inclusion of this coefficient makes the model non-local in space and
requires the introduction of a pressure-like nonlocal term into the
equation. The generalized model allows one to study the effects of pressure
and includes, as special cases, three models that are interesting on their
own: the kinematic MHD model with ${\cal A}=1$ (where the pressure effects
disappear), the linearized Navier--Stokes equation with ${\cal A}=-1$, and
the passive vector ``admixture'' with ${\cal A}=0$~\cite{amodel}--\cite{HA}.

(2) Second, we focus on the effects of strong anisotropy and choose the
Gaussian velocity ensemble as follows: the velocity field is oriented
along a fixed direction ${\bf n}$ (``orientation of a large-scale flare''
in the context of the solar corona dynamics) and depends only on the
coordinates in the subspace orthogonal to ${\bf n}$. In the momentum space,
its correlation function is chosen in the form:
$\langle vv \rangle \propto \delta(t-t')\,  k_{\bot}^{-d+1-\xi}$, where
$k_{\bot}= |{\bf k}_{\bot}|$ and ${\bf k}_{\bot}$ is the component of the
momentum (wave number) ${\bf k}$ perpendicular to ${\bf n}$. This model can
be viewed as
a $d$-dimensional generalization of the strongly anisotropic velocity
ensemble introduced in~\cite{AM} in connection with the turbulent diffusion
problem and further studied and generalized in a number of papers
\cite{AM1}--\cite{AntMal2011}. The model is {\it strongly} anisotropic
in the sense that, in contrast to previous RG+OPE studies of anisotropic
passive advection~\cite{Uni}--\cite{Uni3}, it does not include parameters
that could be tuned to make the velocity statistics isotropic, and hence
it does not include the isotropic Kraichnan's model as a special case.

The problem of anomalous scaling in the higher-order correlation functions of
a {\it scalar} field, advected by such a velocity ensemble, was studied, by
the RG+OPE techniques, in Ref.~\cite{AntMal2011}. It was shown that, in sharp
contrast to the isotropic Kraichnan's model and its numerous descendants, the
correlation functions show no anomalous scaling and have finite limits when
the integral turbulence scale tends to infinity. It should be stressed, that
such a simple behavior has a rather exotic origin: it results from mixing of
{\it families} of relevant composite operators, responsible for the IR
behavior of a given correlation function.
One can say that for typical models the ``normal'' behavior is what
is normally called the ``anomalous'' one.

The main result of the present paper is that the inertial-range behavior
of vector fields advected by such an ensemble is even more exotic: instead
of power-like anomalies, there are logarithmic corrections to ordinary
scaling, determined by naive (canonical) dimensions. The key point is that
the matrices of scaling dimensions (``critical dimensions'' in the
terminology of the theory of critical state) of the relevant families
of composite operators appear nilpotent and cannot be diagonalized.
They can only be brought to Jordan form; hence the logarithms.

It should be stressed that huge families of mixing composite operators are
not unfrequent in field theoretic models, see, e.g., Ref.~\cite{Matraz},
where a set of 3718 operators was encountered in a model of passive vector
advection. But usually the corresponding matrices,
although not symmetric, appear diagonalizable and have real eigenvalues.
The exceptions are known but rare: some models of dense polymers, sandpiles,
dimers and percolation; see Refs.~\cite{Logs} and references therein.
Furthermore, as a rule, the logarithmic behavior is postulated considerately
without a definite Lagrangean field theoretic model, as a hypothetical
continuum limit of discrete evolution models. The model presented in our
paper provides an example of a renormalizable field theoretic model,
where the existence of logarithmic corrections can be proven exactly:
although the formulation of the model is rather cumbersome, it turns out
that the IR behavior is determined completely by the one-loop approximation
of the renormalization group.

To avoid possible misunderstanding, it should be stressed that our ``large
infrared logarithms'' have little to do with the ``large ultraviolet
logarithms,'' known for the $\phi^{4}$ model, quantum electrodynamics, and
in models
of strong interactions (all in $d=4$). In the model under consideration, the
IR logarithms arise due to a highly nontrivial mixing of the relevant
composite operators.

The paper is organized as follows.

In sec.~\ref{sec:Model} we give a detailed description of the model.

In sec.~\ref{sec:QFT} we present the field theoretic formulation
of the model and the corresponding diagrammatic techniques.

In sec.~\ref{sec:Reno} we establish renormalizability of the
(properly extended) model
and derive explicit exact expressions for the renormalization constants
and RG functions (anomalous dimensions and $\beta$ functions).
It is crucial here that the linear response function, the only Green
function in the model that contains superficial UV divergences,
is given exactly by the one-loop approximation.

The RG equations are derived. It is shown that, in some range of
model parameters, they possess an IR attractive fixed point
that governs the IR asymptotic behavior of the correlation functions.
The corresponding differential equations of IR scaling are derived,
with the exactly known critical dimensions.

In sec.~\ref{sec:Ops1} the families of composite operators that give
the leading contributions in the OPE are identified and their
renormalization is discussed. It is shown that the corresponding
renormalization matrices are given exactly by the one-loop approximation.
Explicit expressions for the matrices of
renormalization, anomalous dimensions, and critical dimensions, are
presented. It turns out that the matrices of critical dimensions
cannot be diagonalized. They can be brought to Jordan form with known
diagonal elements. As a result, the dependence of the operator mean values
on the integral turbulence scale is given by known powers, corrected by
polynomials of logarithms.

In sec.~\ref{sec:G-As}, the IR behavior of the pair correlation functions
of the composite operators is discussed. The problem is that, since the
matrices of critical dimensions cannot be diagonalized, those correlation
functions are described by sets of coupled (``entangled'') differential
equations. As a result, their dependence of the separation also involves
polynomials of logarithms.

Eventually, in sec.~\ref{sec:OPE} the solutions of the RG equations for the
mean values and correlation functions of the operators are combined with the
corresponding OPE's to give resulting expressions for the inertial-range
asymptotic behavior of the pair correlation functions. They involve two
types of large logarithms, where the separation enters with the typical
ultraviolet and infrared scales (dissipation scale and integral scale).

Sec.~\ref{sec:Conc} is reserved for conclusions.

Appendix A contains a detailed discussion of the possibility to introduce
two independent length scales for the directions parallel and orthogonal to
the vector ${\bf n}$. Such a possibility exists for the scalar version of
the model, but in our case the transversality condition for the vector
field imposes additional restriction and makes it impossible to have
two scales. However, the problem can be modified such that independent
scales can indeed be introduced.

Appendix B contains a detailed discussion of the derivation of the propagator
matrix in the field theoretic formulation of our model. The subtlety is that,
for transverse vector fields, the standard definition of the propagator as
the matrix inverse to the kernel in the quadratic part of the action
functional does not apply, similarly, e.g., to quantum electrodynamics in the
Loren(t)z gauge. The kernel is not invertible in the full
momentum space and should be inverted
on the transverse subspace. To justify this recipe, we consider a simplified
``toy'' model of a constant random vector field, and give two different
derivations of the propagator.

Appendix C contains the proof of the fact
that the matrices of critical dimensions for all the relevant families of
composite operators are nilpotent. Explicit expressions are presented for
the matrices that bring them to Jordan form. Although the proof looks
rather technical, the statement plays the central role in our analysis of
the IR behavior, and we decided to include it in the full form.

\section{Description of the model} \label{sec:Model}

The turbulent advection of a passive scalar field $\theta(x)\equiv
\theta(t,{\bf x})$ is described by the stochastic equation
\begin{equation}
\label{Kraich-1}
\nabla_{t}\theta =\nu _0\partial^{2} \theta+f, \quad
\nabla_{t} \equiv \partial _t + v_{i}\partial_{i},
\end{equation}
where $\theta(x)$ is the scalar field,
$x\equiv\left\{t,{\bf x}\right\}$, $\partial _t \equiv \partial /\partial t$,
$\partial _i \equiv \partial /\partial x_{i}$, $\nu _0$
is the molecular diffusivity coefficient, $\partial^{2}$ is the Laplace
operator, $\boldsymbol{v}(x)\equiv \{v_{i}(x)\}$ is the transverse (owing
to the incompressibility) velocity field, and $f\equiv f(x)$ is an
artificial Gaussian scalar noise with zero mean and correlation function
\begin{equation}
\label{Kraich-2}
\langle  f(x)  f(x')\rangle = \delta(t-t')\, C({\bf r}/L), \quad
{\bf r}={\bf x}-{\bf x'}.
\end{equation}
The parameter $L$ is an integral scale related to the noise, and
$C({\bf r}/L)$ is some function decaying for $L\to\infty$.

In more realistic formulations, the field $\boldsymbol{v}(x)$ satisfies the
Navier--Stokes (NS) equation. In the rapid-change model it obeys a
Gaussian distribution with zero mean and correlation function
\begin{widetext}
\begin{equation}
\label{Kraich-3}
\langle v_{i}(x) v_{j}(x')\rangle = \delta(t-t')\, D_{0}\
\int_{k>m} \frac{d{\bf k}}{(2\pi)^{d}} \, P_{ij}({\bf k}) \,
 \frac{1}{k^{d+\xi}}\ e^{{\rm i}{\bf k}({\bf x}-{\bf x'})},
\end{equation}
\end{widetext}
where $P_{ij}({\bf k}) = \delta _{ij} - k_i k_j / k^2$ is the transverse
projector, $k\equiv |{\bf k}|$, $D_{0}>0$ is an amplitude factor, $d$ is the
dimensionality of the ${\bf x}$ space and $0<\xi<2$ is a parameter with the
real (``Kolmogorov'') value $\xi=4/3$.

The problem formulated in equations~(\ref{Kraich-1})--(\ref{Kraich-3})
allows for some modifications and generalizations to more complex physical
situations. For example, {\it scalar} diffusion equation~(\ref{Kraich-1})
can be changed to the {\it vector} kinematic MHD
equation, describing, for example, the evolution of the fluctuating part
${\boldsymbol \theta}\equiv {\boldsymbol \theta}(x)$ of the magnetic field
in the presence of a mean component ${\boldsymbol \theta}^o$, which is
supposed to be varying on a very large scale:
\begin{equation}
\label{MHD-Eq}
\partial_t \theta_i+\partial_k\left(v_k\theta_i-\ v_i\theta_k\right)=
\nu_0 \partial^{2} \theta_i+f_i,
\end{equation}
where both $\boldsymbol{v}$ and ${\boldsymbol \theta}$ are divergence-free
(``solenoidal'') vector fields:
\begin{equation}
\label{Trans-Def}
\partial_{i}v_{i}=0, \quad \partial_{i}\theta_{i}=0.
\end{equation}
The linearization of the Navier-Stokes equation
around the rapid-change background velocity field gives the same
expression with a different sign in the vertex term:
\begin{equation}
\label{NS-Eq}
\partial_t \theta_i+\partial_k\left(v_k\theta_i+\ v_i\theta_k\right)+
\partial_{i} {\cal P}=\nu_0 \partial^{2}\theta_i+f_i.
\end{equation}
The pressure term $\partial {\cal P}$ is needed to make the dynamics~(\ref{NS-Eq})
consistent with the transversality conditions
$\partial_{i}\theta_{i}=0$ and $\partial_{i}v_{i}=0$.

Both the equations~(\ref{MHD-Eq}) and~(\ref{NS-Eq}) can be unified by
introducing of a new parameter denoted by ${\cal A}_0$:
\begin{equation}
\label{stoch}
\partial_t \theta_i+\partial_k\left(v_k\theta_i-{\cal A}_0
\ v_i\theta_k\right)+\partial_{i} {\cal P}=\nu_0 \partial^{2}\theta_i+f_i.
\end{equation}

Another interesting case is provided by the choice ${\cal A}_0=0$.
Without the stretching term $\partial_k(v_i\theta_k)$  the model
acquires additional symmetry under translations
${\boldsymbol \theta}\to{\boldsymbol \theta+const}$.
This case has to be studied separately, see Ref.~\cite{Matraz}.

The new parameter requires a new renormalization
constant $Z_{{\cal A}}$, which can be nontrivial
\cite{Ant-Hnat-Hon-Jut-11, AntGul2013}.
The pressure can be expressed as the solution of the Poisson equation
\begin{equation}
\partial^{2} {\cal P} = ({\cal A}_{0} - 1) \,
\partial_{i} v_{k}  \partial_{k} \theta_{i},
\label{Pois}
\end{equation}
so that it vanishes for the local magnetic case.

Of course, when we choose the advection equation like~(\ref{MHD-Eq}),
(\ref{NS-Eq}), or~(\ref{stoch}), we have to modify the correlation
function~(\ref{Kraich-2}). The random external force ${\bf f}$
in the right hand side (RHS) of the equations also becomes a vector,
its statistics is also assumed to be Gaussian,
with zero mean and prescribed correlation function of the form
\begin{equation}
\label{Cik}
\left\langle f_i(t,\ {\bf x})\ f_k(t',\ {\bf x'})\right\rangle
=\delta(t-t')\ C_{ik}({\bf r}/L).
\end{equation}
Like in equation~\eqref{Kraich-2}, here ${\bf r=x-x'}$,
$r=\left|{\bf r}\right|$, the parameter $L\equiv M^{-1}$ is the
integral (external) turbulence scale related to the stirring, and $C_{ik}$
is a dimensionless function finite for $r/L\to0$ and
rapidly decaying for $r/L\to\infty$.

In the real problem, the velocity field  $\boldsymbol{v}(x)$
satisfies the NS equation, probably with additional terms that describe
the feedback of the advected field $\boldsymbol{\theta}(x)$ on the velocity
field. The framework of many works is the {\it kinematic} problem, where
the reaction of the field $\boldsymbol{\theta}(x)$ on the velocity field
$\boldsymbol{v}(x)$ is neglected. It is assumed that, if the gradients of
$\boldsymbol{\theta}(x)$ are not too large, it does not affect essentially
dynamics of the conducting fluid. In this case the latter can be simulated
by statistical ensemble with prescribed statistics.

Here we choose the field $\boldsymbol{v}$ to be strongly anisotropic, namely
having a preferred direction ${\bf n}$:
\begin{equation}
\label{V-n}
\boldsymbol{v}(t,{\bf x})={\bf n}\cdot v(t,\ {\bf x_\perp}).
\end{equation}
It is assumed to be Gaussian, strongly anisotropic (see~(\ref{V-n})),
homogeneous, white-in-time, with zero mean and a correlation function
\begin{equation}
\label{ViVk}
\left\langle v_i(t,\ {\bf x})\ v_k(t',\ {\bf x'})\right\rangle
=n_in_k\ \left\langle v(t,\ {\bf x_\perp})\ v(t',\ {\bf x'_\perp})
\right\rangle,
\end{equation}
where
\begin{equation}
\label{VV-delta}
\left\langle v(t,\ {\bf x_\perp})\ v(t',\ {\bf x'_\perp})\right\rangle
= \delta(t-t')\ \int_{k>m}\frac{d{\bf k}}{(2\pi)^d}\ e^{i{\bf k(x-x')}}\
D_v(k)
\end{equation}
with some function $D_v(k)$, for which we choose
\begin{equation}
\label{D-v}
D_v(k)=2\pi\delta(k_\parallel)\ D_0\ \frac{1}{k_\perp^{d-1+\xi}}.
\end{equation}
Like in equation~\eqref{Kraich-3}, here $d$ is the dimensionality of the
${\bf x}$ space,
$k_\perp \equiv |{\bf k_\perp}|$, $1/m$ is another integral turbulence scale,
related to the stirring,
the exponent $\xi$ plays the role of the RG expansion parameter,
$D_0>0$ is an amplitude factor and symbol $k_\parallel$ denotes
the scalar product $({\bf k\cdot n})$.
The power law~(\ref{D-v}) is suggested
by the experimental data of the turbulence spectra.

To summarize, we will consider the anisotropic vector model,
described by the equations~(\ref{stoch})--(\ref{D-v}).

However, for renormalizability reasons these equations should be generalized
by introducing
one new dimensionless constant $f_0$, which breaks the $O_d$ symmetry
of the Laplace operator to
$O_{d-1}\otimes Z_2$: $\partial^2\rightarrow\
\boldsymbol{\partial}_\perp^2+f_0\partial_\parallel^2$
($Z_{2}$ is the reflection symmetry $x_{\parallel} \to - x_{\parallel}$).
Interpretation of the splitting of the Laplacian term can
be twofold; cf.~\cite{AntMal2011}.
On one hand, stochastic models of the type~(\ref{stoch}) are
phenomenological and, by construction, they must include all
the IR relevant terms allowed by the symmetry. The fact that the splitting is
required by the renormalization procedure means that it is not forbidden by
dimensionality or symmetry considerations and, therefore, it is natural to
include the general value $f_{0}\ne1$ to the model from the very beginning.
On the other hand, one can insist on studying the original model with
$f_{0} =1$ and $O_{d}$ covariant Laplacian term, although that symmetry is
broken to $O_{d-1}\otimes Z_{2}$ by the interaction with the anisotropic
velocity ensemble. 
Then the extension of the model to the case $f_{0}\ne1$ can be
viewed as a purely technical trick which is only needed to ensure the
multiplicative renormalizability and to derive the RG equations. The latter
should then be solved with the special initial data corresponding to
$f_{0} =1$. Since the IR attractive fixed point
of the RG equations is unique (see section~\ref{sec:FP}), the resulting
IR behaviour will be the same as for the general case of the extended
model with $f_{0}\ne1$.

Then the stochastic equation~(\ref{stoch}) takes on the form
\begin{equation}
\label{Stochastic}
\partial_t \theta_i+\partial_k\left(v_k\theta_i-{\cal A}_0\ v_i
\theta_k\right)+\partial {\cal P}=\nu_0\
(\boldsymbol{\partial}_\perp^2+f_0\partial_\parallel^2)\theta_i+f_i.
\end{equation}
The relations
\begin{equation}
\label{D0}
D_0/\nu_0 f_0=\tilde{g}_0\equiv\Lambda^\xi
\end{equation}
define the coupling constant $\tilde{g}_0$, which plays the role of the
expansion parameter in the ordinary perturbation theory, and the
characteristic ultraviolet (UV) momentum scale $\Lambda$.

This completes formulation of the model.

\section{Field theoretic formulation of the model}\label{sec:QFT}

The stochastic problem~(\ref{Cik})--(\ref{Stochastic}) is equivalent to the
field theoretic model of the set of three fields
$\Phi\equiv\left\{\boldsymbol{\theta},\boldsymbol{\theta}',
\boldsymbol{v}\right\}$ with the action functional
\begin{widetext}
\begin{equation}
\label{Action}
{\cal S}(\Phi)= -\frac{1}{2}v_iD_v^{-1}v_k+
\frac{1}{2}\theta'_i D_{\theta}\theta'_k
+\theta'_k\left[-\partial_t\theta_k-(v_i\partial_i)\theta_k
+{\cal A}_0(\theta_i\partial_i)v_k+\nu_0(\partial_\perp^2
+f_0\partial_\parallel^2)\theta_k\right] .
\end{equation}
\end{widetext}
Here all the terms, with the exception of the first one, represent the
De Dominicis--Janssen action for the stochastic
problem~(\ref{Cik}),~(\ref{Stochastic}) at fixed $\boldsymbol{v}$,
while the first term
represents the Gaussian averaging over $\boldsymbol{v}$. Furthermore,
$D_{\theta}$
and $D_{v}$ are the correlators~(\ref{Cik}) and~(\ref{ViVk}) respectively,
the needed integrations over $x=(t,{\bf x})$ and summations over
the vector indices are implied.

The formulation~(\ref{Action}) means that statistical averages of random
quantities in the stochastic problem~(\ref{ViVk}),~(\ref{Stochastic})
coincide with functional averages with weight $\exp {\cal S}(\Phi)$.
The generating functional of the normalized full Green functions
$G = \langle\Phi \dots \Phi \rangle$ is given by the expression
\begin{equation}
\label{Delta-Green}
G(\widetilde{A})=C\cdot\int D\Phi\, \delta\!\left(\partial\theta\right)
\delta\!\left(\partial\theta'\right)e^{{\cal S}(\Phi)+\widetilde{A}\Phi},
\end{equation}
where the normalization constant $C$ is chosen such that $G(0)=1$ and
$\widetilde{A}(x)=\left\{A(x), A'(x), A_v(x)\right\}$ is the set of
``sources,'' arbitrary functional arguments of the same nature as
the corresponding fields.
The presense of two functional $\delta$-functions in the
expression~\eqref{Delta-Green} is a consequence of the second equality
in the expression~\eqref{Trans-Def} and the fact that the auxiliary
field $\boldsymbol{\theta}'$ is also transverse;
see, e.g.,~\cite{Vasiliev-Green}.

The Green functions with the auxiliary field $\boldsymbol{\theta}'$
represent, in the field theoretic formulation, the response functions of
the original stochastic problem, in particular,
the simplest (linear) response function is given by the relation
\begin{equation}
\left\langle \delta\theta_\beta/\delta f_\alpha\right\rangle
= \left\langle\theta_\beta \theta'_\alpha\right\rangle.
\end{equation}

The generating functional of the connected Green functions is given by
\begin{equation}
W(\widetilde{A})=\ln G(\widetilde{A}),
\end{equation}
and the generating functional of the 1-irreducible Green
functions is obtained using the Legendre transform:
\begin{equation}
\label{Legendre}
\Gamma(\Phi)=W(\widetilde{A})-\Phi\widetilde{A},
\end{equation}
where for the functional arguments we have used the same symbols
$\Phi=\left\{\boldsymbol{\theta},\boldsymbol{\theta}',
\boldsymbol{v}\right\}$ as for the corresponding random fields.

\begin{figure}[h!]
\center
\includegraphics[width=.35\textwidth,clip]{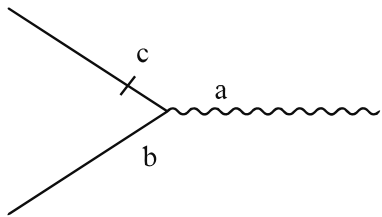}
\caption{The triple vertex.}
\label{fig:TriVert}
\end{figure}

The model~(\ref{Action}) corresponds to a standard Feynman
diagrammatic technique with the triple vertex
$\theta'\left[-(v_i\partial_i)\theta_k+{\cal A}_0(\theta_i\partial_i)
v_k\right]$
and the three bare propagators:
$\left\langle \theta_i\theta'_k\right\rangle_0$,
$\left\langle \theta_i\theta_k\right\rangle_0$ and
$\left\langle v_iv_k\right\rangle_0$ (the propagator
$\left\langle \theta'_i\theta'_k\right\rangle$ is absent).
In the frequency-momentum representation the triple vertex is
written as
\begin{equation}
\label{Triple-vertex}
V_{c\,ab} = i\delta_{bc}\ k_a^{\theta'}
-i{\cal A}_0\delta_{ac}\ k_b^{\theta'},
\end{equation}
where $k^{\theta'}$ is the momentum of the field $\theta'$. In the
diagrammatic notation the vertex is represented in fig.~(\ref{fig:TriVert}).

Strictly speaking, expression~(\ref{Triple-vertex}) should be contracted with
the three transverse projectors for each of the momentum arguments of the
transverse fields $\boldsymbol{v}$, $\boldsymbol{\theta}$
and $\boldsymbol{\theta}'$ entering into the vertex.
However, those projectors will effectively be restored in the diagrams,
when the vertex $V_{c\,ab}$ is contracted with the transverse
propagators (see expressions~\eqref{ThThAux-Propgr}
and~\eqref{ThTh-Propgr} below), so that they can be omitted in
(\ref{Triple-vertex}).
The only exceptions are external vertices in 1-irreducible (amputated)
diagrams, not contracted with external ``legs.'' For these, transverse
projectors with
the corresponding external momenta should be added explicitly.

The three aforementioned propagators are determined by the quadratic
(free) part of the action functional.
They are represented in the diagrams as slashed straight, straight
and wavy lines, respectively:

\begin{figure}[h!]
\center
\includegraphics[width=.28\textwidth,clip]{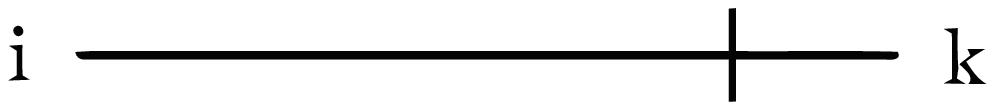}
\caption{Diagrammatic notation for
$\left\langle \theta_i\theta'_k\right\rangle_0$.}
\end{figure}
\begin{figure}[h!]
\center
\includegraphics[width=.28\textwidth,clip]{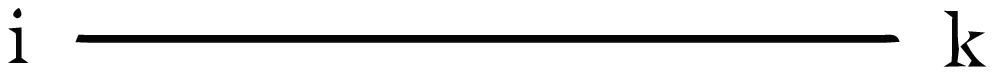}
\caption{Diagrammatic notation for
$\left\langle \theta_i\theta_k\right\rangle_0$.}
\end{figure}
\begin{figure}[h!]
\center
\includegraphics[width=.28\textwidth,clip]{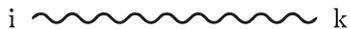}
\caption{Diagrammatic notation for $\left\langle v_iv_k\right\rangle_0$.}
\end{figure}
\noindent

Here the slashed end corresponds to the field $\theta'$,
the end without a slash corresponds to the field $\theta$.
The line $\left\langle v_iv_k\right\rangle_0$ in the diagrams corresponds
to the correlation function~(\ref{ViVk}), and the other two
propagators in the frequency-momentum representation have the forms
\begin{eqnarray}
\label{ThThAux-Propgr}
\left\langle \theta_i\theta'_k\right\rangle_0 &=&
\frac{P_{ik}({\bf k})}{-i\omega+\nu_0({\bf k}_\perp^2+f_0 k_\parallel^2)},
\\
\label{ThTh-Propgr}
\left\langle \theta_i\theta_k\right\rangle_0 &=&
\frac{C_{ik}({\bf k})}{\omega^2+\left[\nu_0({\bf k}_\perp^2
+f_0 k_\parallel^2)\right]^2},
\end{eqnarray}
where $C_{ik}({\bf k})\propto P_{ik}({\bf k})$ is the Fourier transform
of the function from (\ref{Cik}).

In fact, the action functional (\ref{Action}) will be modified for
renormalizability reasons. As a consequence, the functions
(\ref{ThThAux-Propgr}) and~(\ref{ThTh-Propgr}) will acquire certain
additional terms. However, it turns out that those additional
terms do not contribute to the divergent parts of all the relevant
diagrams, and thus can be neglected. These issues are discussed in
detail in sec.~\ref{sec:TruePropagator}, and in the following we will
use for the propagators
the above expressions (\ref{ThThAux-Propgr}) and~(\ref{ThTh-Propgr}).

In the time-momentum representation they take on the forms
\begin{equation}
\label{ThetaFun-Propgr}
\left\langle \theta_i(t) \theta_k'(t') \right\rangle _0 =
P_{ij}({\bf k})\cdot\Theta(t-t')
\exp \left\{ - (t-t')\epsilon_{{\bf k}} \right\},
\end{equation}
\begin{equation}
\left\langle \theta_i(t) \theta_k(t') \right\rangle _0
= \left\{C_{ij}({\bf k})/2\epsilon_{{\bf k}}\right\}\,
\cdot\exp \left\{ - |t-t'| \,
\epsilon_{{\bf k}} \right\},
\end{equation}
where $\epsilon_{{\bf k}}= \nu_0({\bf k}_\perp^2+f_0k_\parallel^2)$.
The propagator $\langle \theta_i \theta_k' \rangle _0$ is retarded.

\section{From renormalization to critical dimensions}
\label{sec:Reno}

\subsection{Canonical dimensions and UV divergences} \label{sec:Canon}

The analysis of UV divergences is based on the analysis of canonical
dimensions of the 1-irreducible Green functions. In general,
dynamic models have two scales: canonical dimension of some
quantity $F$ (a field or a parameter in the action functional) is completely
characterized by two numbers, the frequency dimension $d_{F}^{\omega}$
and the momentum dimension $d_{F}^{k}$. They are defined such that
\begin{equation}
\label{F-Propto-Dim}
[F] \sim [T]^{-d_{F}^{\omega}} [L]^{-d_{F}^{k}},
\end{equation}
where $L$ is some reference length scale and $T$ is a time scale.

In the {\it scalar} version of our strongly anisotropic model
\cite{AntMal2011}, however, there are two independent length
scales, related to the directions perpendicular and parallel to the
vector ${\bf n}$: namely, one can introduce two independent momentum
canonical dimensions $d_{F}^{\bot}$ and $d_{F}^{\parallel}$ so that
$[F] \sim [T]^{-d_{F}^{\omega}}  [L_{\bot}]^{-d_{F}^{\bot}}
[L_{\parallel}]^{-d_{F}^{\parallel}}$, where $L_{\bot}$ and
$L_{\parallel}$ are (independent) length scales in the corresponding
subspaces. In the present vector model, however, we have an
additional condition of the transversality of the fields $\theta$
and $\theta'$:
\begin{equation}
\label{transvers}
\partial_i\theta_i=0, \quad \partial_i\theta'_i=0,
\end{equation}
which prevents the introduction of two independent
scales. This issue is discussed in Appendix~\ref{App:III} in detail.
In particular, this means that, in contrast to the scalar case, the
constant $f_0$  from~(\ref{Stochastic}) in our case is dimensionless.

The dimensions in (\ref{F-Propto-Dim}) are found from the obvious
normalization conditions
$d_{k}^{k}=-d_{\bf x}^{k}=1$,
$d_{k}^{\omega}= -d_{\bf x}^{\omega}=0$,
$d_{\omega}^{\omega} = -d_{t}^{\omega}=1$,
$d_{\omega }^{k}=d_t^{k}=0$,
and from the
requirement that each term of the action functional~(\ref{Action})
be dimensionless (with respect to the two independent dimensions
separately).
Based on $d_{F}^{k}$ and $d_{F}^{\omega}$, one can introduce the
total canonical dimension $d_{F}=d_{F}^{k}+2d_{F}^{\omega}$
(in the free theory,
$\partial_{t}\propto\partial^{2}_{\bot} \propto \partial^{2}_{\parallel}$),
which plays in the theory of renormalization of dynamic models the same
role as the conventional (momentum) dimension does in static problems;
see, e.g.,~\cite{Vasiliev-Green}.

\begin{table*}
\caption{Canonical dimensions of the fields and parameters
in the model (\protect\ref{Action})}
\begin{ruledtabular}
\label{table1}
\begin{tabular}{c||c|c|c|c|c|c|c|c|c|c}
$F$ & $\theta' $ & $\theta$ & $ \boldsymbol{v} $ &  $M,m,\mu, \Lambda $ &
$\nu ,\nu_{0}$ & ${\cal A} ,{\cal A}_{0}$ & $f, f_{0}$ & $u, u_0$
&  $\tilde{g}_{0}$, $g_{0}$ & $\tilde{g}$, $g$ \\
\hline
$d_{F}^{\omega}$ & 1/2 & $-1/2$ & 1 & 0 & 1 & 0 & 0 & 0 & 0&  0\\
\hline
$d_{F}^{k}$ & $d$ & 0 & $-1$ & 1 & $-2$ & 0 & 0 & 0 & $\xi$ & 0 \\
\hline
$d_{F}$ & $d+1$ & $-1$ & 1 & 1 & 0 & 0 & 0 & 0 &
$\xi$ & 0 \\
\end{tabular}
\end{ruledtabular}
\end{table*}

The canonical dimensions for the model~(\ref{Action}) are given in
Table~\ref{table1}, including renormalized parameters, which will be
introduced a bit later. From Table~\ref{table1} it follows that our model
is logarithmic (the coupling constant $g_{0} \sim [L]^{-\xi}$
is dimensionless) at $\xi=0$, so that the UV divergences manifest
themselves as poles in $\xi$ in the Green functions.

The total canonical dimension of an arbitrary 1-irreducible Green function
$\Gamma_{N_\Phi} = \langle\Phi \dots \Phi \rangle _{\rm 1-ir}$ is given by
the relation
\begin{equation}
d_{\Gamma_{N_\Phi}}= d+2- \sum_{\Phi} N_{\Phi }d_{\Phi} = d+2-
N_{\theta'} d_{\theta'} - N_{\theta} d_{\theta} - N_{v} d_{v}.
\label{dGamma}
\end{equation}
Here $N_{\Phi}=\{N_{\theta},\,N_{\theta'},\,N_{v}\}$ are the numbers of
corresponding fields entering the function $\Gamma_{N_\Phi}$, and the
summation
over all types of the fields in~(\ref{dGamma}) and analogous formulae below
is always implied.

Superficial UV divergences, whose removal requires counterterms, can be
present only in those functions $\Gamma_{N_\Phi}$ for which the
``formal index of
divergence'' $d_{\Gamma_{N_\Phi}}$ is a non-negative integer.
Dimensional analysis should be augmented by the following observations:

(1) In any dynamical model of type~(\ref{Action}), 1-irreducible
diagrams with $N_{\theta'}=0$ contain closed circuits of retarded
propagators~(\ref{ThetaFun-Propgr}) and therefore vanish.

(2) For any 1-irreducible Green function $N_{\theta'}- N_{\theta}=2N_{0}$,
where $N_{0}\ge0$ is the total number of the bare propagators
$\langle \theta \theta \rangle _0$ entering into any of its diagrams. This
fact is easily checked for any given function; it is illustrated by the
function with $N_{\theta'}= N_{\theta}=1$ and $N_{0}=0$,
see fig.~\ref{fig:SelfEnergy}. Clearly, no diagrams with $N_{0}<0$ can
be drawn. Therefore, the difference $N_{\theta'}- N_{\theta}$ is an
even non-negative integer for any nonvanishing function.

(3) Using the transversality condition of the fields $\theta$ and $v$
we can move one derivative from the vertex
$-\theta'_k(v_i\partial_i)\theta_k+{\cal A}_0\
\theta'_k(\theta_i\partial_i)v_k$
onto the field $\theta'_k$.
Therefore, in any
1-irreducible diagram it is always possible to move the derivative onto
external ``tail'' $\theta'_k$, which reduces the real
index of divergence: $d_{\Gamma_{N_\Phi}}' = d_{\Gamma_{N_\Phi}}-N_{\theta'}$.
The field $\theta'_k$ enters into the counterterms only in the
form of derivative $\partial_{i}\theta'_k$.

From Table~\ref{table1} and~(\ref{dGamma}) we find:
\begin{equation}
d_{\Gamma_{N_\Phi}}= d+2 - (d+1) N_{\theta'} + N_{\theta} -N_{v}
\end{equation}
and
\begin{equation}
d'_{\Gamma_{N_\Phi}}\! =(d+2)(1-N_{\theta'}) + N_{\theta} - N_{v}.
\end{equation}
From these expressions we conclude that, for any $d$, superficial
divergences can be present only in the 1-irreducible functions of two types.

The first example is provided by the infinite family of functions
$\langle\theta'\theta\dots\theta\rangle_{\rm 1-ir}$ with $N_{\theta'}=1$
and arbitrary $N_{\theta}$, for which $d_{\Gamma}=2$, $d_{\Gamma}'=0$.
However, all the functions with $N_{\theta}\geq N_{\theta'}$ vanish
(see above) and obviously do not require counterterms. Therefore the
only nonvanishing function from this family is
$\langle\theta_\alpha'\theta_\beta\rangle_{\rm 1-ir}$.

Another possibility is
$\langle \theta'\theta\dots\theta v\dots v\rangle_{\text{1-ir}}$ with
$N_{\theta'}=1$ and arbitrary $N_{\theta}=N_{v}$, for
which $d_{\Gamma}=1$, $d_{\Gamma}'=0$.
From the requirement $N_{\theta}\geq N_{\theta'}$ it follows
that the nonvanishing function of this type is
$\langle \theta'_\alpha\theta_\beta v_\gamma\rangle_{\text{1-ir}}$.
Furthermore, from the explicit expressions
\eqref{VV-delta} and~(\ref{ThetaFun-Propgr}) for the propagators
it follows, that all the diagrams for that function contain closed circuits
of retarded lines and therefore vanish.

Thus we are left with the only superficially divergent function
$\langle\theta_\alpha'\theta_\beta\rangle_{\rm 1-ir}$.


\subsection{Perturbation expansion for the 1-irreducible linear response
function}

\begin{figure}[h!]
\center
\includegraphics[width=.35\textwidth,clip]{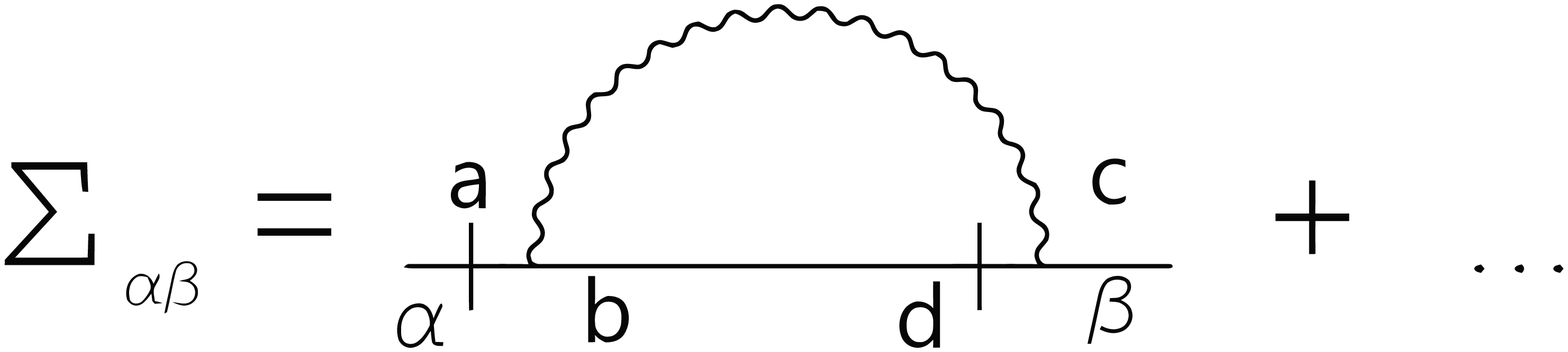}
\caption{Diagrammatic representation for $\Sigma_{\alpha\beta}$.}
\label{fig:SelfEnergy}
\end{figure}

Consider the 1-irreducible linear response function
\begin{equation}
\left.\Gamma_2^{\alpha\beta}=\langle\theta_\alpha'\theta_\beta
\rangle_{\rm 1-ir}=
\frac{\delta}{\delta \theta'_\alpha}\frac{\delta}{\delta \theta_\beta}
\Gamma(\Phi)\right|_{\Phi=0},
\label{Vasik2}
\end{equation}
where the generating function of the 1-irreducible Green
functions $\Gamma(\Phi)$ (see~\eqref{Legendre}) consists of two parts,
\begin{equation}
\Gamma(\Phi) =  {\cal S}(\Phi) + \widetilde{\Gamma}(\Phi).
\end{equation}
Here ${\cal S}(\Phi)$ is the action functional (\ref{Action}) and
$\widetilde{\Gamma}(\Phi)$ is the sum of all the 1-irreducible diagrams
with loops. Due to transversality of the fields in (\ref{Vasik2}), the
result of formal differentiation  should be contracted with transverse
projectors. Thus for the function $\Gamma_2$ one obtains
\begin{widetext}
\begin{equation}
\label{Dyson}
\Gamma_2^{\alpha\beta}=i\omega\cdot P_{\alpha\beta}({\bf p})-\nu_0
{\bf p}_\perp^2\cdot P_{\alpha\beta}({\bf p})-\nu_0f_0\cdot({\bf pn})^2
\cdot P_{\alpha\beta}({\bf p})+\Sigma_{\alpha\beta},
\end{equation}
\end{widetext}
where $P_{\alpha\beta}({\bf p})=\delta_{\alpha\beta}-p_\alpha p_\beta/p^2$
is transverse projector and $\Sigma_{\alpha\beta}$ is the
``self-energy operator,''
diagrammatic representation for which is represented in the
fig.~\ref{fig:SelfEnergy}.
\noindent
Here the ellipsis stands for the 2-, 3- and other N-loop diagrams.

The typical feature of all the rapid-change models
(\ref{VV-delta}) with retarded bare propagator~(\ref{ThetaFun-Propgr})
is that all the skeleton multiloop diagrams entering into the self-energy
operator contain closed circuits of such retarded propagators
and therefore vanish.
Thus the self-energy operator in~(\ref{Dyson}) is {\it exactly} given by
the one-loop approximation.

Let us begin the calculation of the diagram with its index structure:
\begin{equation}
\label{Trans-J}
J_{\alpha\beta}=
P_{\alpha i}({\bf p})\cdot J_{ij}\cdot P_{j\beta}({\bf p}),
\end{equation}
where
\begin{equation}
J_{ij}= V_{i\, ab}({\bf p})V_{d\,cj}({\bf p+k})P_{bd}({\bf p+k})n_an_c.
\end{equation}
Here $V_{ijk}({\bf p})$ is the triple vertex~(\ref{Triple-vertex});
the Greek letters $\alpha$, $\beta$ and the Roman letters $a$--$d$
denote the vector indices of the propagators~(\ref{ViVk}),
(\ref{ThThAux-Propgr}),~(\ref{ThTh-Propgr}) with the implied summation
over repeated indices. Note that we need to calculate only the divergent
part of the diagram, i.e., only the terms, proportional
to ${\bf p}^2$. Some observations simplify the calculation:

(1) Because of our choice of $D_v$ (see~(\ref{D-v})),
namely its proportionality to $\delta(k_\parallel)$, all the terms
proportional to $k_\parallel$
vanish after the integration over the momentum ${\bf k}$.

(2) Both the main field $\boldsymbol{\theta}$ and the auxiliary
field $\boldsymbol{\theta'}$ are divergence-free:
$\partial_\beta\theta_\beta=p_\beta\theta_\beta=0$;
$\partial_\alpha\theta'_\alpha=p_\alpha\theta'_\alpha=0$. Thus, all the
terms proportional to $p_\alpha$ or $p_\beta$
disappear after the contraction with the external fields
$\boldsymbol{\theta}$ and $\boldsymbol{\theta}'$
(see the remarks below expression~\eqref{Triple-vertex}
and the expression~\eqref{Trans-J}).

This gives the following expression for the index structure of
$\Sigma_{\alpha\beta}$-diagram:
\begin{widetext}
\begin{equation}
\label{J12}
J_{ij} = -\delta_{ij}
\cdot({\bf pn})^2 - ({\cal A}_0-1)\cdot({\bf pn})^2\cdot
\frac{k_i k_j}{k^2} + {\cal A}_0({\cal A}_0-1)
\cdot({\bf pn})\cdot\frac{({\bf pk})k_j n_i}{k^2},
\end{equation}
where we retained the terms of order ${\bf p}^{2}$.

Now we have to integrate this expression over the $d$-dimensional momentum
${\bf k}$ with the factor $1/(2\pi)^d\cdot D_v(k)$ with $D_v(k)$ from
(\ref{D-v}) and over the frequency $\omega$ with the factors $1/2\pi$ and
$1/(-i\omega+\nu_0[({\bf p+k})_\perp^2+f_0(p+k)_\parallel^2])$:
\begin{equation}
\label{Sigma-General-Expression}
\Sigma_{\alpha\beta}=\int\frac{d\omega}{2\pi}\int\frac{d{\bf k}}
{(2\pi)^d}\frac{2\pi\, D_0\,  \delta(k_\parallel)\, D_0}
{(-i\omega+\nu_0[({\bf p+k})_\perp^2+f_0(p+k)_\parallel^2])}\cdot
P_{\alpha i}({\bf p}) J_{ij} P_{j\beta}({\bf p})/k_\perp^{d-1+\xi}.
\end{equation}
\end{widetext}
The integration over the frequency $\omega$ is simple due to the following
interpretation of the Heaviside step function at coincided times:
\begin{equation}
\Theta(t-t')={1}/{2}
\end{equation}
at $t=t'$, which is justified by the fact that the correlation function
is symmetric in its arguments; cf.~\cite{RG}. Thus
\begin{equation}
\label{Omega-Extending-Definition}
\int\frac{d\omega}{2\pi}\frac{1}{(-i\omega+\nu_0[({\bf p+k})_
\perp^2+f_0(p+k)_\parallel^2])}=\frac{1}{2}.
\end{equation}

The integration over ${\bf k}$ with the function $\delta(k_\parallel)$
in the integrand is performed with the aid of the relations
\begin{eqnarray}
\label{Int-Aver-Angles}
\int d{\bf k}\delta(k_\parallel) f({\bf k})=
\int d{\bf k}_{\perp}\, f({\bf k}_{\perp})=
\nonumber \\
= S_{d-1} \int_m^\infty dk_{\perp}\, k_{\perp}^{d-1}\,
\left\langle f({\bf k}_{\perp})\right\rangle,
\end{eqnarray}
where $\langle\cdots\rangle$ is the averaging over the unit sphere in the
$(d-1)$-dimensional space, $S_{d-1}$ is its surface area, and
$k_{\perp}= |{\bf k}_{\perp}|$.

Now we need to average the two structures $k_i k_j/k^2$
and $({\bf pk}) k_j/k^2$ from eq.~(\ref{J12}) with the replacement
$k_{i}\to k_{i}^{{\perp}}$. The first structure is orthogonal to
${\bf n}$, so the result of averaging is proportional to the
transverse projector $P_{ij}({\bf n})=\delta_{ij}-n_in_j$:
\begin{equation}
\label{k-transv-Aver-Angles}
\left\langle \frac{k_i^\perp k_j^\perp}{k_\perp^2} \right\rangle
= \frac{P_{ij}({\bf n})}{(d-1)},
\end{equation}
where the coefficient is found by comparing the traces of the right
and left hand sides.

Multiplying~\eqref{k-transv-Aver-Angles} to $p_i$, for the second structure
we obtain
\begin{equation}
\label{pkCdotK-Aver-Angles}
\left\langle \frac{({\bf pk}_{\perp})k^{\perp}_j}{k^2_\perp}\right\rangle
= \frac{p_i P_{ij}({\bf n})}{(d-1)} \simeq
 -\frac{({\bf pn})n_j}{(d-1)}.
\end{equation}
In the last relation we omitted a term proportional to $p_{j}$ because it
will vanish after the contraction with the transverse projector
$P_{j\beta}({\bf p})$ in expression (\ref{Sigma-General-Expression}).

After the angular averaging have been performed, we are left with the simple
integral over the modulus $k_\perp$:
\begin{equation}
\int_m^\infty \frac{dk_\perp}{k_\perp^{1+\xi}} = \frac{m^{-\xi}}{\xi}.
\end{equation}
Combining the expressions
(\ref{J12}),~(\ref{Int-Aver-Angles}),~(\ref{k-transv-Aver-Angles})
and~(\ref{pkCdotK-Aver-Angles}) we obtain for
(\ref{Sigma-General-Expression}) the following result:
\begin{widetext}
\begin{equation}
\label{Sigma-Answer}
\Sigma_{\alpha\beta} = -\frac{1}{2}\cdot D_0\cdot C_{d-1}\cdot
\left[\frac{d-2+{\cal A}_0}{d-1}
\cdot P_{\alpha\beta}({\bf p})+\frac{({\cal A}_0-1)^2}{d-1}\cdot
\hat{n}_\alpha \hat{n}_\beta\right]\cdot({\bf pn})^2\cdot
\frac{m^{-\xi}}{\xi}.
\end{equation}
\end{widetext}
Here $C_{d-1}\equiv S_{d-1}/(2\pi)^{d-1}$, the amplitude $D_0$ was
introduced in~(\ref{D0}), and the vector $\hat{n}_k$,  defined as
\begin{equation}
\hat{n}_k = P_{mk}({\bf p})n_m = n_k-p_\parallel p_k/p^2,
\end{equation}
is orthogonal to ${\bf p}$.


\subsection{Renormalization and bare propagators}
\label{sec:TruePropagator}

Substituting the explicit expression~(\ref{Sigma-Answer}) for the
divergent part of the self-energy operator $\Sigma_{\alpha\beta}$
to the expression~(\ref{Dyson}) for the 1-irreducible linear response
function $\Gamma_2^{\alpha\beta}$ gives
\begin{widetext}
\begin{equation}
\label{Dyson-2}
\Gamma_2^{\alpha\beta}= \{ i\omega\cdot -\nu_0 {\bf p}_\perp^2
-\nu_0f_0\cdot({\bf pn})^2 \} \cdot P_{\alpha\beta}({\bf p})
-D_0\cdot\left[\frac{d-2+{\cal A}_0}{2(d-1)}
\cdot P_{\alpha\beta}({\bf p})+\frac{({\cal A}_0-1)^2}{2(d-1)}\cdot
\hat{n}_\alpha \hat{n}_\beta\right]\cdot C_{d-1}\cdot({\bf pn})^2\cdot
\frac{m^{-\xi}}{\xi}.
\end{equation}
\end{widetext}

The renormalization constants are found from the requirement that the
function (\ref{Dyson-2}), when expressed in new renormalized variables,
be UV finite, i.e., finite at $\xi\to0$. From the analysis of this
expression it follows, however, that the pole in $\xi$ in the structure
with $\hat{n}_\alpha \hat{n}_\beta$ cannot be removed by renormalization
of the model parameters, because the bare part of $\Gamma_2^{\alpha\beta}$
does not contain analogous term. In order to ensure multiplicative
renormalizability one has to add such term, with a new positive amplitude
factor $u_{0}$, to the bare part:
\begin{widetext}
\begin{eqnarray}
\Gamma_2^{\alpha\beta} &=& \left\{i\omega
-\nu_0{\bf p}_\perp^2
-\nu_0f_0\cdot({\bf pn})^2\right\}\cdot P_{\alpha\beta}({\bf p})
-\nu_0f_0u_0\cdot({\bf pn})^2\cdot
\hat{n}_\alpha \hat{n}_\beta-
\nonumber \\
&-& D_0\cdot\left[\frac{d-2+{\cal A}_0}{2(d-1)}
\cdot P_{\alpha\beta}({\bf p})+\frac{({\cal A}_0-1)^2}{2(d-1)}\cdot
\hat{n}_\alpha \hat{n}_\beta\right]\cdot C_{d-1}\cdot({\bf pn})^2\cdot
\frac{m^{-\xi}}{\xi}.
\label{Dyson-True}
\end{eqnarray}
\end{widetext}
This means that the original model (\ref{Action}) is extended by adding the
term of the form
$u_{0}f_{0}\nu_0 (n_k\theta_k') \partial_\parallel^2\cdot (n_k\theta_k)$;
the interpretation of the new parameter $u_{0}$ is literally the same as
for $f_{0}$ in sec.~\ref{sec:Model}.

Now the poles can be eliminated by multiplicative renormalization of
the parameters $f_0$, $u_0$, and $g_0$:
\begin{equation}
\label{Renorm-Parameters}
f_0=f Z_{f}, \ u_0=u Z_{u}, \ g_{0}=g\mu^{\xi}Z_{g}, \ Z_{g}=Z_{f}^{-1}.
\end{equation}
Here $\mu$ is the ``reference mass'' (additional free parameter of the
renormalized theory) in the minimal subtraction (MS) renormalization
scheme, which we always use in what follows, $g$, $u$ and $f$
are renormalized analogs of the bare parameters $g_{0}$, $u_0$ and $f_0$,
and $Z_i=Z_i(g,\xi,d)$ are the renormalization constants. Their
relation in~(\ref{Renorm-Parameters}) results from the absence of
renormalization of the contribution with $D_v^{-1}$ in~(\ref{Action}),
so that $D_{0}\equiv g_{0}\nu_0f_0 = g\mu^{\xi} \nu f$.

No renormalization of the fields and the parameters $\nu_0$, ${\cal A}_0$
and $m_{0}=m$ is required: i.e., $Z_{\Phi}=1$ for all $\Phi$ and
$Z_{\nu,m,{\cal A}}=1$. Explicit expressions for nontrivial renormalization
constants will be given in the next subsection.

The renormalized action functional has the form
\begin{widetext}
\begin{eqnarray}
\label{RenAction}
{\cal S}_R(\Phi) &=& \frac{1}{2}\theta'_iD_{\theta}\theta'_k
-\frac{1}{2}v_iD_v^{-1}v_k+
\theta'_k\left[-\partial_t\theta_k-(v_i\partial_i)\theta_k+{\cal A}
(\theta_i\partial_i)v_k+\nu(\partial_\perp^2+fZ_f\cdot \partial_\parallel^2)
\theta_k\right]+
\nonumber \\
&+& \nu\cdot fZ_f\cdot uZ_u\cdot (n_k\theta_k')
\partial_\parallel^2\cdot(n_k\theta_k),
\end{eqnarray}
\end{widetext}
where the function $D_v$ from~(\ref{D-v}) is expressed in renormalized
parameters using~(\ref{Renorm-Parameters}).

One important question arises here. Since the original model was
modified by adding a new term to the action~(\ref{Action}), the old
expressions for the propagators \eqref{ThThAux-Propgr} and
\eqref{ThTh-Propgr} have also changed. One may think that the whole
calculation of the self-energy operator, performed in the preceding
subsection, should be repeated with the new propagators. Below we show
that additional terms in the modified bare propagator do not contribute
to the integral~(\ref{Sigma-General-Expression}) and revision of the
final expression~(\ref{Dyson-True}) is in fact not needed.

Let us denote the bare contribution in~(\ref{Dyson-True}) as
$-M_{\alpha\beta}$, so that
\begin{equation}
\label{Propogrt-X-Y-Def}
M_{\alpha\beta} =
{\cal X}\cdot P_{ij}({\bf p}) + {\cal Y} \cdot \hat{n}_i\hat{n}_j,
\end{equation}
where
\begin{eqnarray}
{\cal X} &=&-i\omega+\nu_0{\bf p}_\perp^2+\nu_0f_0\cdot({\bf pn})^2,
\label{X} \\
{\cal Y} &=& \nu_0f_0u_0\cdot({\bf pn})^2.
\label{Y}
\end{eqnarray}
In a conventional situation, the bare propagator
$\left\langle \theta\theta'\right\rangle_0$ would be given by the inverse
matrix $M_{\alpha\beta}^{-1}$. Our matrix (\ref{Propogrt-X-Y-Def}) has a
nontrivial eigenvector with zero eigenvalue, $M_{\alpha\beta}p_{\beta}=0$,
so the inverse matrix does not exist. However, the functional integration
in (\ref{Delta-Green}) is taken over the subspace of transverse fields.
Thus we do not need to invert the matrix (\ref{Propogrt-X-Y-Def}) on the
full momentum space, rather we need to invert it on the subspace orthogonal
to the vector ${\bf p}$, where the role of the unity matrix is played by the
transverse projector. Therefore the desidered ``inverse'' matrix
\[ M^{-1}_{\alpha\beta}=  x\cdot P_{ij}({\bf p})
+ y \cdot \hat{n}_i\hat{n}_j\]
is found from the equation
\begin{equation}
\label{Definite-N}
M_{\alpha\beta}({\bf p})\cdot M_{\beta\gamma}^{-1}({\bf p}) =
P_{\alpha\gamma}({\bf p}).
\end{equation}
This gives
\begin{subequations}
 \begin{equation}
  x=1/{\cal X};
 \end{equation}
 \begin{equation}
  \label{Propogtr-Y}
  y=-{\cal Y}/{\cal X}({\cal X}+{\cal Y}\sin^2\kappa),
 \end{equation}
\label{xy}
\end{subequations}
where $\kappa$ is the angle between the vectors ${\bf n}$ and ${\bf p}$.

Thus for the propagator matrix in the modified model we obtain
\begin{equation}
\label{Propgtr-Full-WithPart}
\left\langle \theta_{\alpha}\theta'_{\beta}\right\rangle_0
= x\cdot P_{\alpha\beta}({\bf p})
+ y \cdot \hat{n}_\alpha\hat{n}_\beta
\end{equation}
with the scalar coefficients $x$ and $y$ from eq.~(\ref{xy}).
The first term coincides with the propagator \eqref{ThThAux-Propgr}
of the original model (\ref{Action}) up to the notation.

From the expressions~(\ref{Propogrt-X-Y-Def}) and (\ref{Propogtr-Y}) for $y$
as a function of $\omega$ one can write
\begin{equation}
\label{Ptopgtr-Y-Propto}
y=-{\cal Y}/{\cal X}({\cal X}+{\cal Y}\sin^2\kappa)\propto
\frac{({\bf pn})^2}{(-i\omega+\eta_1)(-i\omega+\eta_2)}
\end{equation}
with certain $\eta_1$, $\eta_2 >0$ that depend only on the momentum.
The both poles in $\omega$ in expression~(\ref{Ptopgtr-Y-Propto}) lie
in the same lower half of the complex plane. Thus the integral over
$\omega$ of this expression vanishes identically. It is important here that
the integral is convergent by power counting, and the ambiguity similar
to that in the integral~(\ref{Omega-Extending-Definition}) is absent here.

This means that the new term in the propagator does not contribute
to the integral~(\ref{Sigma-General-Expression}) and the
expression~(\ref{Dyson-True}) remains valid in the modified model.
Of course, the contribution with $y$ shoud be taken into account in
calculations of the other Green functions.

Some additional remarks about our scheme of derivation of the propagators
and its justification on the example of a simple model
are given in Appendix~\ref{Sec:Toymodel}.

\subsection{RG equations and fixed points} \label{sec:FP}

Now let us introduce the $\beta$ functions and the anomalous dimensions
$\gamma$~-- important RG functions, which determine the asymptotic
behavior of the various Green functions. The basic RG equation for a
multiplicatively renormalizable quantity (correlation function,
composite operator, etc.) is obtained by operating with
$\widetilde{{\cal D}}_\mu$ on the relation $F=Z_{F}F_{R}$, where
$\widetilde{\cal D}_{\mu}$ denotes the differential operation
$\mu\partial_{\mu}$ for fixed set of bare parameters
$e_{0}$ $=$ $\left\{g_{0}, \nu_{0}, f_{0}, u_{0}, {\cal A}_{0}\right\}$.
The resulting RG equation has the form
\begin{equation}
\label{RG-Eqtn}
\bigl[{\cal D}_{RG}+ \gamma_{F} \bigr] F_{R}=0,
\end{equation}
where $\gamma_F$ is the anomalous dimension of $F$ and
${\cal D}_{RG} = {\cal D}_{\mu} + \beta \partial_{g} -
\gamma_{f}{\cal D}_{f}- \gamma_{u}{\cal D}_{u}$.
Here and below ${\cal D}_{x}\equiv x\partial_{x}$ for any variable $x$,
and the RG functions are defined as
\begin{subequations}
\begin{equation}
\label{Beta-g-Def}
\beta_g \equiv \widetilde{\cal D}_\mu g =g\cdot[-\xi-\gamma_{g}(g)],
\end{equation}
\begin{equation}
\label{Beta-u-Def}
\beta_u \equiv \widetilde{\cal D}_\mu u =-u\gamma_{u}(g,u),
\end{equation}
\begin{equation}
\gamma_{F} \equiv \widetilde{\cal D}_\mu \ln Z_{F} =
\beta_g \partial_{g} \ln Z_{F} \quad  {\rm for\ any\ } Z_{F}.
\end{equation}
\end{subequations}
The relations between $\beta$ and $\gamma$ in~(\ref{Beta-g-Def}) and
(\ref{Beta-u-Def})
result from their definitions along with the second and third relations in
(\ref{Renorm-Parameters}).

As already stated, the constants $Z$ are found from the requirement of UV
finiteness of the expression~(\ref{Dyson-True}) by means of the relations
$D_0 g_{0}f_{0}\nu_0=gf\nu\mu^{\xi}$ and~(\ref{Renorm-Parameters}). Thus
we readily obtain the renormalization constant
$Z_f$ ($f_0=f\cdot Z_f$) and the anomalous dimension $\gamma_f$ for
the parameter $f_0$ that splits the Laplace operator:
\begin{equation}
Z_f=1-\frac{d-2+{\cal A}}{2(d-1)}\cdot\frac{g}{\xi}+O(g^2),
\end{equation}
\begin{equation}
\label{gamma-f}
\gamma_f=\frac{d-2+{\cal A}}{2(d-1)}\cdot g,
\end{equation}
where we passed to the new coupling constant $g\equiv\tilde{g}\cdot C_{d-1}$
with $C_{d-1}$ from (\ref{Sigma-Answer}).
Then we have to renormalize the constant $u_0$ such that the expression
\begin{equation}
g_0f_0u_0\cdot\left[1+\frac{({\cal A}-1)^2}{2(d-1)}
\cdot\frac{1}{u_0}\cdot\frac{m^{-\xi}}{\xi}\right]\cdot
n_\alpha n_\beta\cdot({\bf pn})^2
\end{equation}
be UV finite to the first order in $g$. Therefore,
\begin{equation}
Z_u\cdot Z_f=1-\frac{({\cal A}-1)^2}
{2(d-1)}\cdot\frac{g}{u}\cdot\frac{1}{\xi}+O(g^2)
\end{equation}
and
\begin{equation}
\gamma_u+\gamma_f=\frac{({\cal A}-1)^2}{2(d-1)}\cdot\frac{1}{u}\cdot g
\end{equation}
with the constant $\gamma_f$ from~\ref{gamma-f}.
Furthermore, from the last relation in~(\ref{Renorm-Parameters}) it follows
that for the coupling constant $g$
\begin{equation}
\label{gamma-g-gamma-f}
\gamma_g=-\gamma_f= -\frac{d-2+{\cal A}}{2(d-1)}\cdot g.
\end{equation}

We stress that all the above expressions for the anomalous dimensions
$\gamma_{F}$ are exact, being derived from the exact
expression~(\ref{Dyson-True}).

One of the basic RG statements is that the IR asymptotic behavior of the
model is governed by the IR attractive fixed point $g^*$, $u^*$, defined by
the relations
\begin{equation}
\beta_u=0, \ \ \partial_u\beta_u>0;\ \ \beta_g=0, \ \ \partial_g\beta_g>0.
\end{equation}
For the coupling constant $g$ these equations along
with~\eqref{gamma-g-gamma-f} give
\begin{equation}
\beta_g=g(-\xi+\gamma_f)=0,
\end{equation}
and the fixed point is
\begin{equation}
\label{g-fixed}
g^*=\frac{2(d-1)}{d-2+{\cal A}}\cdot\xi, \ \ \partial_g\beta_g(g^*)=\xi>0.
\end{equation}

The $\beta$-function and the fixed point for second parameter $u$ are
\begin{equation}
\beta_u=-u\gamma_u=g\cdot\frac{1}{2(d-1)}\left[(d-2+{\cal A})
\cdot u-({\cal A}-1)^2\right],
\end{equation}
so that
\begin{equation}
u^*=\frac{({\cal A}-1)^2}{d-2+{\cal A}},
\ \ \partial_u\beta_u(u^*)=\frac{d-2+{\cal A}}{2(d-1)}\cdot g^*.
\end{equation}
Therefore, the system possesses an IR fixed point $u^*, g^*$ only if $g^*>0$,
i.e.,
\begin{equation}
d-2+{\cal A}>0.
\end{equation}

This fact implies that the correlation functions of the model~(\ref{Action})
in the IR region ($\mu r\simeq\Lambda r \gg 1$, $Mr \sim 1$) exhibit scaling
behavior (as we will see below, up to logarithmic factors).

The corresponding critical dimensions $\Delta[F]\equiv\Delta_{F}$ for all
basic fields and parameters can be calculated exactly; see the next
subsection.

\subsection{Critical dimensions}

In the leading order of the IR asymptotic behavior the Green functions
satisfy the RG equation with the substitution $g\to g_{*}$, $u\to u_{*}$,
which gives
\begin{equation}
\left[ {\cal D}_{\mu} - \gamma_{f}^{*}{\cal D}_{f} -
\gamma_{u}^{*}{\cal D}_{u}
+ \gamma_{G}^{*} \right] \,G^{R}(e,\mu,\dots) = 0.
\label{RGFP}
\end{equation}
Canonical scale invariance is expressed by the relations:

\begin{equation}
\left[\sum _{\alpha}d_{\alpha}^k{\cal D}_{\alpha}-
d_{G}^k\right]G^{R}=0 ,\quad
\left[\sum _{\alpha}d_{\alpha}^{\omega }{\cal D}_{\alpha}-
d_{G}^{\omega }\right]G^{R}=0 ,
\label{Canonic-Scl-Inv}
\end{equation}
where $\alpha\equiv\{t,{\bf x},\mu,\nu,m,M,u,f,{\cal A},g\}$ is the set of
all arguments of $G^{R}$ ($t,{\bf x}$ is the set of all times
and coordinates), and $d^{k}$ and $d^{\omega}$ are the
canonical dimensions of $G^{R}$ and $\alpha$. Substituting the
needed dimensions from Table~\ref{table1} into~(\ref{Canonic-Scl-Inv}),
we obtain:

\begin{subequations}
\label{Scaling-Equations}
\begin{equation}
\left[{\cal D}_{\mu}+{\cal D}_{m}+{\cal D}_{M}-2{\cal D}_{\nu}
-{\cal D}_{\bf x}-d_{G}^{k}\right]G^{R} = 0,
\end{equation}

\begin{equation}
\left[{\cal D}_{\nu}-{\cal D}_{t}-d_{G}^{\omega}\right]G^{R} = 0.
\end{equation}
\end{subequations}

The equations of the type~(\ref{RGFP}) and~(\ref{Scaling-Equations})
describe the scaling behavior of the function $G^{R}$ upon the dilation
of a part of its parameters. A parameter is dilated if the corresponding
derivative enters the equation, otherwise it is kept fixed. We are interested
in the IR scaling behavior, in which all the IR relevant parameters
(coordinates ${\bf x}$, times $t$ and integral scales $M$ and $m$)
are dilated, while the irrelevant parameters, related to the UV scale
(diffusivity coefficient $\nu$ and the renormalization mass $\mu$) are fixed.
Thus we combine the equations~(\ref{RGFP}) and~(\ref{Scaling-Equations})
such that the
derivatives with respect to the IR irrelevant parameters $\mu$
and $\nu$ be eliminated, and obtain
the desired equation of critical IR scaling for the model:

$$\left[-{\cal D}_{\bf x}+ \Delta_{t} {\cal D}_{t} +
\Delta_{m} {\cal D}_{m} + \Delta_{M} {\cal D}_{M} +\right.$$

\begin{equation}
\label{IR-Scaling}
\left.+ \Delta_{f} {\cal D}_{f} + \Delta_{u} {\cal D}_{u} -
\Delta_{G} \right]G^{R} = 0,
\end{equation}
where
$$\Delta_{t}=-\Delta_{\omega}=-2, \quad
\Delta_{m} = \Delta_{M} =1, $$
\begin{equation}
\label{IR-Scaling-Coeff}
\Delta_{f} =\gamma_f^*,  \quad
\Delta_{u} =\gamma_u^*
\end{equation}
and
\begin{equation}
\Delta[G]\equiv\Delta_{G} = d_{G}^{k}+ 2d_{G}^{\omega}+\gamma_{G}^{*}
\label{Critical-Dim}
\end{equation}
are the corresponding critical dimensions.

In particular, for any correlation function
$G^{R}=\langle \Phi\dots\Phi\rangle$ of the fields $\Phi$
we have $\Delta_{G} = N_{\Phi} \Delta_{\Phi}$, with the summation
over all fields $\Phi$ entering into $ G^{R}$, namely

\begin{equation}
\Delta_{G}= \sum_{\Phi} N_{\Phi}d_{\Phi} = N_{\theta'}d_{\theta'}+
N_{\theta}d_{\theta}+ N_{v}d_{v}.
\label{Ca}
\end{equation}
Since in the model~(\ref{Action}) the fields themselves are not renormalized
(i.e., $\gamma_{\Phi}=0$ for all $\Phi$,
see sec.~\ref{sec:TruePropagator}), using~(\ref{Critical-Dim}) we conclude,
that the critical dimensions of the fields $\Phi=\left\{\boldsymbol{v},
{\boldsymbol \theta},{\boldsymbol \theta'}\right\}$ are the same as their
canonical dimensions, presented in the Table~\ref{table1}. Namely,

\begin{equation}
\Delta_{\boldsymbol{v}}=1,\quad
\Delta_{\theta} = -1,\quad
\Delta_{\theta'} = d+1.
\label{33}
\end{equation}
It is the specific feature of this model, which distinguishes it from
both the isotropic Kraichnan's vector model~\cite{AntGul2012}
(in which $\gamma_\nu\neq0$) and anisotropic Kraichnan's scalar model
\cite{AntMal2011} (in which the Laplacian splitting parameter $f$ is
not dimensionless).

\section{Renormalization and critical dimensions
of composite operators} \label{sec:Ops1}

\subsection{General scheme}
\label{sec:scheme}

In the following, the central role will be played by composite fields
(``operators'') built solely of the basic fields $\theta$:
\begin{equation}
\label{F-N-p}
F_{Np}=(\theta_i\theta_i)^p\ (n_s\theta_s)^{2m},
\end{equation}
where $N=2(p+m)$ is the total number of fields $\theta$, entering
the operator.

The total canonical dimension of arbitrary 1-irreducible Green function
$\Gamma = \langle F \Phi\dots\Phi\rangle_{1-ir}$
that includes one composite operator $F$ and arbitrary number of
primary fields $\Phi$ is given by the relation
\begin{equation}
\label{D-Comp}
d_{\Gamma}=d_F-\sum_\phi N_\Phi d_\Phi = -(N-N_\theta)
-(d+1)N_{\theta'}-N_v,
\end{equation}
where $d_F$ is the canonical dimension of the operator $F$; see, e.g.,
sec.~3.24 in the book \cite{Vasiliev-Green}.

Superficial UV divergences, whose removal requires counterterms, can be
present only in those functions $\Gamma$ for which the index of
divergence $d_{\Gamma_{N_\Phi}}$ is a non-negative integer.
For the operators of the form~\eqref{F-N-p} one has $d_F=-N$;
see Table~\ref{table1}. Due to the linearity of our model,
the number of the fields $\theta$ in the function $\Gamma$ cannot exceed
their number in the operator $F$; cf.~item~(2) in sec.~\ref{sec:Canon}.
In the case at hand
\begin{equation}
\label{D-Comp-2}
N_\theta\leq N.
\end{equation}
From the expression~\eqref{D-Comp} along with the condition~\eqref{D-Comp-2}
it follows that for any dimension of the space $d$ superficial divergences
can be present only in the 1-irreducible functions of the type
$\bigl\langle F_{Np} (x) \theta(x_{1})\dots\theta(x_{N})
\bigr\rangle_{\rm 1-ir}$ with $N_\theta=N$, $N_v=N_{\theta'}=0$,
for which $d_{\Gamma_F}=0$. This means, in particular, that
all those diagrams diverge logarithmically and we can calculate them
with all external frequencies and momenta set equal to zero. This also
means that the operator counterterms to a certain $F_{Np}$ include
only operators of the form~\eqref{F-N-p} with the same value of $N$.

We conclude that the operators~\eqref{F-N-p} can mix in renormalization
only within the closed set with the same $N$; let us denote it as
$F\equiv\{F_{Np}\}$.
The renormalization matrix $\hat{Z}_{F}\equiv\{Z_{Np,Np'}\}$
for this set, given by the relation
\begin{equation}
F_{Np}=\sum _{p'} Z_{Np,Np'} F_{Np'}^{R},
\end{equation}
is determined by the requirement that the 1-irreducible correlation function
\begin{widetext}
$$\bigl\langle F_{Np}^{R} (x) \theta(x_{1})\dots\theta(x_{N})
\bigr\rangle_{\rm 1-ir}=$$
\begin{equation}
=\sum _{p'} Z_{Np,\,Np'}^{-1}\bigl\langle
F_{Np'}(x) \theta(x_{1})\dots\theta(x_{N})\bigr\rangle
_{\rm 1-ir} \equiv \sum _{p'}Z_{Np,\,Np'}^{-1}\Gamma_{Np'}
(x;x_{1},\dots, x_{N})
\end{equation}
\end{widetext}
be UV finite in renormalized theory, that is, it has no poles in $\xi$ when
expressed in renormalized variables~(\ref{Renorm-Parameters}). This is
equivalent to the UV finiteness of the sum
$\sum_{p'}Z_{Np,\,Np'}^{-1}\Gamma_{Np'}(x;\theta)$, in which
$$\Gamma_{Np'} (x;\theta) = \frac{1}{N!}\, \int d x_{1}\dots \int d x_{N}\,
\Gamma_{Np'} (x;x_{1},\dots, x_{N})\times$$
\begin{equation}
\label{Gamma-Np}
\times\theta(x_{1})\dots\theta(x_{N})
\end{equation}
is a functional of the field $\theta(x)$.

The contribution of a specific diagram into the functional $\Gamma_{Np'}$
in~(\ref{Gamma-Np}) for any composite operator $F_{Np'}$ is represented in
the form
\begin{equation}
\label{Diag-General}
\Gamma_{Np'} = V_{\alpha\beta\dots} \, I^{ab\dots}_{\alpha\beta\dots} \,
\theta_{a} \theta_{b} \dots ,
\end{equation}
where $V_{\alpha\beta\dots}$ is the vertex factor,
$I^{ab\dots}_{\alpha\beta\dots}$ is the ``internal block''  of the diagram
with free indices, and the product $\theta_{a} \theta_{b} \dots$ corresponds
to external tails.

According to the general rules of the universal diagrammatic technique
(see, e.g.,~\cite{Vasiliev-Green}), for any composite operator $F(x)$
built of the fields $\theta$, the vertex $V_{\alpha\beta\dots}$ in~(\ref{Diag-General})
with $k\ge0$ attached lines corresponds to the vertex
factor

\begin{equation}
\label{Vertex-General}
V^{k}_{Np} (x;\, x_{1}, \dots, x_{k}) \equiv \delta^{k}
F_{Np}(x) / {\delta\theta(x_{1}) \dots\delta\theta(x_{k})}.
\end{equation}
The arguments $x_{1}\dots x_{k}$ of the quantity~(\ref{Vertex-General}) are
contracted with the arguments of the upper ends of the tails $\theta\theta'$
attached to the vertex.

\subsection{One-loop diagram}

\begin{figure}[h]
\center
\includegraphics[width=.22\textwidth,clip]{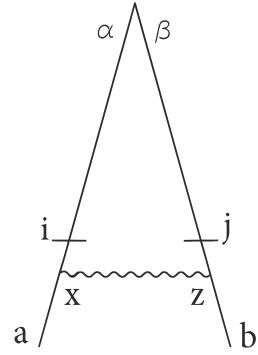}
\caption{The one-loop contribution to the generating functional.}
\label{fig:One-Loop}
\end{figure}

Now let us turn to calculate the internal block
$I^{ab\dots}_{\alpha\beta\dots}$ in the notation~(\ref{Diag-General}),
namely the diagrams themselves.
The one-loop diagram is represented in
fig.~(\ref{fig:One-Loop}).

\noindent

We recall that all the external frequencies and momenta are set to zero.
Then the index structure of this diagram is

$$Y^{ab}_{\alpha\beta} = V_{xai}({\bf k})\ V_{zjb}({\bf -k})
\cdot P_{\alpha i}({\bf k})\ P_{\beta j}({\bf k})\cdot n_x n_z=$$
 \begin{equation}
=-{\cal A}^2\cdot n_xP_{x\alpha}({\bf k})\cdot n_zP_{z\beta}({\bf k})
\cdot k_ak_b,
\end{equation}
where the letters $i,j,x$ and $z$ denote internal indices of the diagram
itself. Then we have to integrate $Y^{ab}_{\alpha\beta}$ over the frequency
and momentum with the factors like~\eqref{VV-delta}
and~(\ref{ThThAux-Propgr}), namely

$$I^{ab}_{\alpha\beta} = \int\frac{d{\bf k}}{(2\pi)^d}\cdot\frac{1}
{-i\omega+\nu {\bf k}_\perp^2+\nu fk_\parallel^2}\cdot\frac{1}{i\omega+
\nu {\bf k}_\perp^2+\nu fk_\parallel^2}\times$$
\begin{equation}
\label{Oper-1Loop-General}
\times D_0\cdot  \frac{\delta(k_\parallel)}{k_\perp^{d-1+\xi}}\cdot
Y^{ab}_{\alpha\beta}.
\end{equation}

Due to the presence of the factor $\delta(k_\parallel)$ in the above
expression, the additional part~\eqref{Ptopgtr-Y-Propto} in the propagator
$\langle\theta\theta'\rangle_{0}$ gives no contribution in all integrals
corresponding to the diagrams, and like for the self-energy function, we
can perform all the  calculations with the original
propagators~\eqref{ThThAux-Propgr} and~\eqref{ThTh-Propgr}.

Using~(\ref{k-transv-Aver-Angles}) for averaging over the angles,
we arrive at the following result:
$$
I^{ab}_{\alpha\beta} = \frac{{\cal A}^2}{2\nu}\cdot D_0\cdot
\int\frac{d{\bf k}_\perp}{(2\pi)^d}\
\frac{1}{k_\perp^{d-1+\xi}}\cdot\frac{k_a^\perp k_b^\perp}{k^2_\perp}\cdot
n_\alpha n_\beta=
$$
\begin{equation}
\label{1loop-diag-answer}
=\frac{{\cal A}^2\cdot f}{2\ (d-1)}\cdot P_{ab}({\bf n})\cdot
n_\alpha n_\beta\cdot\ g\cdot\frac{m^{-\xi}}{\xi}.
\end{equation}

\subsection{Multiloop diagrams}

\begin{figure}[h]
\center
\includegraphics[width=.15\textwidth,clip]{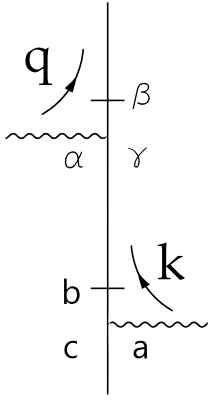}
\caption{Fragment of {\it arbitrary} multiloop diagram.}
\label{fig:Multi-Loop}
\end{figure}

Any multiloop diagram contains a part with the structure,
represented in fig.~(\ref{fig:Multi-Loop}).

\noindent
Since we may calculate all the diagrams at external momenta set equal
to zero, the integral, corresponding to the divergent part of
the diagram, contains as a factor the following expression:
\begin{equation}
I_0=\delta(k_\parallel)\delta(q_\parallel)n_aV_{bac}({\bf k})
n_\alpha V_{\beta\alpha\gamma}({\bf k+q})P_{\gamma b}({\bf k}),
\end{equation}
where $V$ is the vertex~(\ref{Triple-vertex}), and the $\delta$-functions
appear from velocity correlator~(\ref{ViVk}). Since $I_0$ is proportional
to the sum of $k_\parallel$ and $q_\parallel$ with some coefficients, after
integration with the $\delta$-functions all these diagrams become equal to
zero.

There are also multiloop diagrams of the ``sand clock'' type, represented
by products of simpler diagrams. They contain only higher-order poles in
$\xi$ and, in the MS scheme, do not contribute to the anomalous dimensions.

Therefore (and it is another special feature of this model) the
one-loop approximation~(\ref{1loop-diag-answer}) gives us the {\it exact}
answer.


\subsection{Renormalization matrix and anomalous dimensions}
\label{sec:AmonDim}

Combining expressions~(\ref{Diag-General}),~(\ref{Vertex-General}) and
the exact answer~(\ref{1loop-diag-answer}) for the diagram we obtain
for the functional $\Gamma_{Np}$ from (\ref{Gamma-Np})
\begin{widetext}
\begin{eqnarray}
\Gamma_{Np} &\propto&
\frac{\delta^2}{\delta\theta_\alpha\cdot\delta\theta_\beta}
\left[F_{Np}\right]\cdot n_\alpha n_\beta\cdot P_{ab}({\bf n})\
\theta_a\theta_b =
\nonumber \\
&=& 2m(2m-1)\cdot F_{N\,p+1}\ +\ (2p+8pm-2m(2m-1))\cdot F_{N\,p}\ +
\nonumber \\
&+& (4p(p-1)-2p-8pm)\cdot F_{N\,p-1}\ -\ 4p(p-1)\cdot F_{N\,p-2},
\label{F-diag1}
\end{eqnarray}
\end{widetext}
up to an overall scalar factor.

Expression~(\ref{F-diag1}) shows that the operators $F_{Np}$ indeed
mix in renormalization: the UV finite renormalized operator $F^{R}$ has
the form $F^{R}=F+$ counterterms, where the contribution of the
counterterms is a linear combination of $F$ itself and other unrenormalized
operators with the same total number $N$ of the fields, which  are said
to ``admix'' to $F$ in renormalization.

Let $F\equiv\{F_{p}\}$ be a closed set of operators (\ref{F-N-p}) with a
certain fixed value of $N$ (which we will omit below for  brevity) and
different values of $p$, which mix only to each other in renormalization.
The renormalization matrix $\hat{Z}_{F}\equiv\{Z_{p,p'}\}$
and the  matrix of anomalous dimensions
$\hat{\gamma}_{F}\equiv\{\gamma_{p,p'}\}$
for this set are given by
\begin{equation}
\label{gammaF-Def}
F_{p}=\sum _{p'} Z_{p,p'} F_{p'}^{R},
\quad
\hat{\gamma}_F=\hat{Z}_{F}^{-1}{\cal D}_{\mu }\hat{Z}_{F}.
\end{equation}
The scale invariance~(\ref{Canonic-Scl-Inv}) and the RG equation
(\ref{RG-Eqtn})
for the operator $F_{p}$ give us the corresponding matrix of critical
dimensions
$\Delta_{F}\equiv\{\Delta_{p,p'}\}$ in the form similar to the expression
(\ref{Critical-Dim}), in which
$d_{F}^{k}$, $d_{F}^{\omega}$ and $d_{F}$ are understood as the
diagonal matrices of canonical dimensions of the operators in
question (with the diagonal elements equal to sums of corresponding
dimensions of all fields and derivatives constituting $F$) and
$\hat{\gamma}^{*}=\hat{\gamma} (g^{*},u^{*})$ is the matrix
(\ref{gammaF-Def}) at the fixed point.

In this notation and in the MS scheme the renormalization matrix $\hat{Z}$
has the form
\begin{equation}
\label{Z-F-Common}
\hat{Z} = \hat{I} + \hat{A},
\end{equation}
where $\hat{I}$ is the unity matrix and the elements of the matrix
$\hat{A}$ have the form
\begin{equation}
\label{Z-diagram-matrix}
A_{pp'}=a_{pp'}\cdot\frac{g}{\xi}.
\end{equation}

To solve the RG equation we have to use the eigenvalue decomposition of the
matrix $\hat{\gamma}$, therefore the critical dimensions of the set
$F\equiv\{F_{p}\}$ are
given by the eigenvalues of the matrix $\Delta_{pp'}$. In fact this means,
that we change the set of operators $\left\{F^R\right\}$  to the set of
``basis''
operators $\left\{\widetilde{F}^R\right\}$ that possess definite critical
dimensions and have the form

\begin{equation}
\label{FF-tld}
F^R_p=U_{pp'}\widetilde{F}^R_{p'},
\end{equation}
where the matrix $U_{pp'}$
is such that $ \widetilde{\Delta}_{F}= U_{F}^{-1} \Delta_{F} U_{F}$
is diagonal (or has the Jordan form).

As the renormalization matrix $\hat{Z}$ has the form~(\ref{Z-F-Common}),
the matrix of anomalous dimensions $\hat{\gamma}$ has the form
\begin{equation}
\label{Gamma-Common-Aik}
\gamma_{pp'}=-a_{pp'}\cdot g
\end{equation}
with the coefficients $a_{pp'}$ from~(\ref{Z-diagram-matrix}). Combining
(\ref{F-diag1})--(\ref{Gamma-Common-Aik}) and taking into account the
scalar factor, not written in~(\ref{F-diag1}), but presented
in~(\ref{1loop-diag-answer}), together with the fact, that the symmetrical
coefficient for this one-loop diagram is $1/2$, one can obtain the following
for the matrix of anomalous dimensions $\hat{\gamma}$:

\begin{eqnarray}
\label{gamma-F}
\gamma_{p,\,p'+1}&=&-\frac{{\cal A}^2\cdot f}{4(d-1)}\cdot2m(2m-1)
\cdot g;
\nonumber \\
\gamma_{p,\,p'}&=&-\frac{{\cal A}^2\cdot f}{4(d-1)}\cdot(2p+8pm-2m(2m-1))
\cdot g; \nonumber \\
\gamma_{p,\,p'-1}&=&-\frac{{\cal A}^2\cdot f}{4(d-1)}\cdot(4p(p-1)-2p-8pm)
\cdot g; \nonumber \\
\gamma_{p,\,p'-2}&=&-\frac{{\cal A}^2\cdot f}{4(d-1)}\cdot(-4p(p-1))
\cdot g.
\end{eqnarray}

Substituting the value of the fixed point
$g^*=\frac{2(d-1)}{d-2+{\cal A}}\cdot\xi$ (see~(\ref{g-fixed})) gives
\begin{eqnarray}
\label{gamma-F-fixed}
\gamma_{p,\,p'+1}^*&=&-\frac{{\cal A}^2\cdot f}{2(d-2+{\cal A})}
\cdot2m(2m-1)\cdot\xi; \nonumber \\
\gamma_{p,\,p'}^*&=&-\frac{{\cal A}^2\cdot f}{2(d-2+{\cal A})}
\cdot(2p+8pm-2m(2m-1))\cdot\xi; \nonumber \\
\gamma_{p,\,p'-1}^*&=&-\frac{{\cal A}^2\cdot f}{2(d-2+{\cal A})}
\cdot(4p(p-1)-2p-8pm)\cdot\xi; \nonumber \\
\gamma_{p,\,p'-2}^*&=&-\frac{{\cal A}^2\cdot f}{2(d-2+{\cal A})}
\cdot(-4p(p-1))\cdot\xi.
\end{eqnarray}
Therefore the critical dimensions matrix for the set $F_{p}$ has the
form
\begin{equation}
\label{Crit-Dim-F}
\Delta_{p,\,p'}=-2(p+m)\cdot\delta_{pp'}+\hat{\gamma}^*_{p,\,p'},
\end{equation}
where $-2(p+m)$ is its canonical dimension, $\delta_{pp'}$ is Kronecker's
$\delta$ symbol and $\hat{\gamma}^*_{p,p'}$ is the value of anomalous
dimension matrix at the fixed point.


\subsection{Critical dimension matrix and its eigenvalue decomposition}
\label{sec:Diag}

Let us find the eigenvalues of the critical dimensions matrix
$\Delta_{Np,Np'}$ (from now on, we restore the index $N$ in the notation
for the operators and related matrices).
As a consequence of~(\ref{gamma-F}), it is a four-diagonal matrix
for any $N$; moreover it has one line under the main diagonal and two lines
above the main diagonal. Therefore the inversion of the matrix and its
eigenvalue decomposition appear nontrivial tasks.

According to~(\ref{F-diag1}), the closed set $F\equiv\{F_{Np}\}$ of operators,
which mix only with each other in renormalization, consists only of
operators with the same total quantity of fields $\theta$, i.e.,
with the same number $N$. So, let us define the vector ${\bf F}$ as
\begin{equation}
\label{F-Vector-Def}
{\bf F}=
\begin{pmatrix}
(\theta_i\theta_i)^N\\
(\theta_i\theta_i)^{N-2}\cdot(n_s\theta_s)^2\\
\vdots \\
(n_s\theta_s)^N
\end{pmatrix}.
\end{equation}
Therefore the relation between the set of unrenormalized operators
$\left\{F\right\}$ and the set of renormalized operators
$\left\{F^R\right\}$, namely $F_i=Z_{ik}F_k^R$, takes on the form

\begin{widetext}
\begin{equation}
\label{F-Z-Fr}
\begin{pmatrix}
(\theta_i\theta_i)^N\\
(\theta_i\theta_i)^{N-2}\cdot(n_s\theta_s)^2\\
(\theta_i\theta_i)^{N-4}\cdot(n_s\theta_s)^4\\
\vdots \\
\vdots \\
(\theta_i\theta_i)^{2}\cdot(n_s\theta_s)^{N-2}\\
(n_s\theta_s)^N
\end{pmatrix}
=
\begin{pmatrix}
a_{11} & a_{12} & a_{13} & 0 & \dots & 0\\
a_{21} & a_{22} & a_{23} & a_{24} & &  \vdots\\
0 & a_{32} & a_{33} & a_{34} & \ddots & 0\\
\vdots & 0 & a_{43} & \ddots & \ddots & a_{n-2n}\\
\vdots &  &  & \ddots & \ddots & a_{n-1n}\\
0 &\dots & \dots & 0 & a_{nn-1} & a_{nn}
\end{pmatrix}
\cdot
\begin{pmatrix}
\left\{(\theta_i\theta_i)^N\right\}^R\\
\left\{(\theta_i\theta_i)^{N-2}\cdot(n_s\theta_s)^2\right\}^R\\
\left\{(\theta_i\theta_i)^{N-4}\cdot(n_s\theta_s)^4\right\}^R\\
\vdots \\
\vdots \\
\left\{(\theta_i\theta_i)^{2}\cdot(n_s\theta_s)^{N-2}\right\}^R\\
\left\{(n_s\theta_s)^N\right\}^R
\end{pmatrix}.
\end{equation}
\end{widetext}
It is important that in this notation the row of the matrix $\hat{Z}$
corresponds to the original unrenormalized operator, and that the
power $p$ of the operator $F_{Np}$ decreases from left to right.

Let us denote the common factor in~(\ref{gamma-F-fixed}) as $y$, i.e.,

\begin{equation}
y=-\frac{{\cal A}^2\cdot f}{2(d-2+{\cal A})}\cdot\xi,
\end{equation}
and construct from~(\ref{gamma-F-fixed}),~(\ref{Crit-Dim-F}) and
(\ref{F-Z-Fr}) the matrix of critical dimensions for several sets of
operators. For example, for the set with $N=2$ we have

\begin{equation}
N=2:\ \ \
\Delta_{Np,Np'}=
\begin{pmatrix}
-2+2y & -2y \\
2y & -2-2y
\end{pmatrix},
\end{equation}
and the eigenvalues are $\lambda=\left\{-2;-2\right\}$;
for the set with $N=8$ we have

$$N=8:\ \ \
\Delta_{Np,Np'}=
$$

\begin{equation}
=
\begin{pmatrix}
-8+8y & 40y & -48y & 0 & 0 \\
2y & -8+28y & -6y & -24y & 0 \\
0 & 12y & -8+24y & -28y & -8y\\
0 & 0 & 30y & -8-4y  & -26y \\
0 & 0 & 0 & 56y  & -8-56y
\end{pmatrix},
\end{equation}
and the eigenvalues are $\lambda=\left\{-8;-8;-8;-8;-8\right\}$;
and so on. This fact remains true for ${\it any}$ set of operators
with {\it arbitrary} number $N$. This statement is strictly proven in
Appendix~\ref{App:I}.

In other words, for {\it any} $N$, the matrix of anomalous dimensions
(\ref{gamma-F-fixed}) is nilpotent, and the matrix of critical dimensions
(\ref{Crit-Dim-F}) is degenerate with all the eigenvalues equal to $N$:

\begin{equation}
\lambda_1=\ \dots\ =\lambda_{N/2+1}=-2(p+m)=-N.
\end{equation}
Therefore the matrix of critical dimensions~(\ref{Crit-Dim-F}) is not
diagonalizable, but can be brought to a Jordan form, i.e.,
$\Delta_{F}= U_{F} \widetilde{\Delta}_{F} U_{F}^{-1}$, and for the matrix
$\widetilde{\Delta}_{F}$ we can write

\begin{equation}
\label{Delta-F-Tilde}
\widetilde{\Delta}_{F}
=
\begin{pmatrix}
-2(p+m) & 1 & 0 & \dots & 0 \\
0 & -2(p+m) & 1 &  & \vdots \\
\vdots & 0 & \ddots & \ddots  &0 \\
\vdots &  &  & \ddots & 1 \\
0 & \dots &  & 0 & -2(p+m)
\end{pmatrix}.
\end{equation}
The matrix $U_F$, which brings it to the Jordan form, is triangular, namely

\begin{equation}
\label{U-UnDef}
U_{F}
=
\begin{pmatrix}
u_{11} & u_{12} & u_{13} & \dots & \dots & u_{1n}\\
u_{21} & u_{22} & \dots & \dots & u_{2n-1} &  0\\
u_{31} & \vdots &  &~\reflectbox{$\ddots$} &  & \vdots\\
\vdots & \vdots &~\reflectbox{$\ddots$} & \ &  & \vdots\\
u_{n-1\ 1} & u_{n-1\ 2} &  &  &  & \vdots\\
u_{n1} & 0 & \dots & \dots & \dots & 0
\end{pmatrix},
\end{equation}
with the elements $u_{ik}\neq0$ for all $i$, $k$ (for detailed discussion
see Appendix~\ref{App:I}).


\subsection{Asymptotic behavior of the mean value of operator $F_{Np}$}

The objects of interest are, in particular, the equal-time correlation
functions $G=\left\langle F_1F_2 \right\rangle$. In using the operator
product expansion (OPE), the mean values of the operators
$\left\langle F^R \right\rangle$ will appear in the right hand side
(see below). Therefore now it is useful to understand the asymptotic behavior
of the quantities $\left\langle F^R \right\rangle$ themselves.

From the dimensionality considerations it follows that

\begin{equation}
\label{Matrix-tmp4}
\left\langle {\bf \widetilde{F}}^R \right\rangle\propto
\nu^{d^\omega_F}\mu^{d_F}
\cdot \widehat{\Phi}\left(\frac{M}{\mu},f\right)
\cdot\widehat{{\bf C}}_0.
\end{equation}
If an operator $F$ itself is multiplicatively renormalizable,
in the IR-region it satisfies the differential equation
(\ref{IR-Scaling})--(\ref{IR-Scaling-Coeff}), which describes the
IR scaling behavior. The solution of this equation (the mean value does
not depend on the time $t$ and coordinates ${\bf x}$) gives us the
asymptotic form:

\begin{equation}
\left\langle {\bf \widetilde{F}}^R \right\rangle\propto
M^{\widetilde{\Delta}_F}\cdot \Phi\left(\frac{f}{M^{\gamma_f^*}}\right)
\cdot{\bf C}_0.
\end{equation}
This along with the dimensionality representation~\eqref{Matrix-tmp4} gives
\begin{equation}
\label{F-R-RG-Asympt}
\left\langle {\bf \widetilde{F}}^R \right\rangle\propto
\nu^{d^\omega_F}\mu^{d_F}
\cdot\left(M/\mu\right)^{\widetilde{\Delta}_F}
\cdot \Phi\left(\frac{f}{(M/\mu)^{\xi}}\right)\cdot{\bf C}_0,
\end{equation}
where $\Phi$ is some unknown function of the dimensionless argument,
$\gamma_f^*=\xi$ (see equations~(\ref{gamma-f}) and~(\ref{g-fixed})),
${\bf \widetilde{F}}^R$ is a vector built from the ``basis'' operators
(\ref{FF-tld}) that possess definite critical dimensions, ${\bf C}_0$
is some constant vector (``initial data''), $\mu$ is the renormalization mass, $\nu$ is the viscosity coefficient,
$M$ is the parameter, related to the external turbulence scale connected to the stirring (see expression~\eqref{Cik})
%
%
and $\widetilde{\Delta}_F$
is the matrix of critical dimensions from~(\ref{Delta-F-Tilde}). 


Since the matrix $\widetilde{\Delta}_F$ in~(\ref{F-R-RG-Asympt}) has a
Jordan form with the only degenerate eigenvalue $\lambda_0=-2(p+m)$,
then the value of a certain scalar function ${\cal F}$ with the  matrix
argument $\widetilde{\Delta}_F$ is given by the matrix
${\cal F}\left(\widetilde{\Delta}_{F}\right)$:

\begin{equation}
{\cal F}\left(\widetilde{\Delta}_{F}\right)
=
\begin{pmatrix}
{\cal F}\left(\lambda_0\right) & \frac{{\cal F}'\left(\lambda_0\right)}{1!}  & \dots & \frac{{\cal F}^{(n-1)}\left(\lambda_0\right)}{(n-1)!} \\
0 & {\cal F}\left(\lambda_0\right) & & \vdots \\
\vdots &    & \ddots & \frac{{\cal F}'\left(\lambda_0\right)}{1!} \\
0 & \dots  & 0 & {\cal F}\left(\lambda_0\right)
\end{pmatrix}.
\end{equation}
If the function ${\cal F}$ is chosen as $(M/\mu)^{\widetilde{\Delta}_F}$,
the logarithms $\ln (M/\mu)$ will appear in the sought-for asymptotic
expression:

\begin{widetext}
\begin{equation}
\label{MR-Delta}
(M/\mu)^{\widetilde{\Delta}_F}
=
\begin{pmatrix}
(M/\mu)^\lambda & (M/\mu)^\lambda\cdot\ln (M/\mu)  &
\dots & \frac{(M/\mu)^\lambda\cdot(\ln (M/\mu))^{n-1}}{(n-1)!} \\
0 & (M/\mu)^\lambda & & \vdots \\
\vdots &    & \ddots & (M/\mu)^\lambda\cdot\ln (M/\mu) \\
0 & \dots  & 0 & (M/\mu)^\lambda
\end{pmatrix}.
\end{equation}
\end{widetext}
Therefore, after the convolution with the initial-data vector ${\bf C}_0$
and up to the dimensional factor, the asymptotic form of the mean value
of the operators $\widetilde{F}^R$ is

\begin{equation}
\label{F-Tilde-Asympt}
\begin{array}{l}
\widetilde{F}^R_1\propto (M/\mu)^\lambda \cdot P_{N/2}
\left(\ln (M/\mu)\right), \\
\widetilde{F}^R_2\propto (M/\mu)^\lambda \cdot P_{N/2-1}
\left(\ln (M/\mu)\right), \\[8pt]
\vdots \\[8pt]
\widetilde{F}^R_{N/2+1}\propto (M/\mu)^\lambda,
\end{array}
\end{equation}
where $P_L\left(\dots\right)$ is a polynomial of degree $L$
with the argument $\ln (M/\mu)$.
Note that the indices $1,\ \dots,\ N/2+1$ in~(\ref{F-Tilde-Asympt}) are not
``accidental'': they are strongly related with the indices of the vector
${\bf F}$, which is defined in~(\ref{F-Vector-Def}).

\section{Asymptotic behavior of the correlation function
$G=\left\langle F_1F_2 \right\rangle$}
\label{sec:G-As}

Now we are ready to begin studying the IR asymptotic behavior of the
correlation function of the two composite operators
$F_{Np}$ of form~(\ref{F-N-p}) with arbitrary values of $N$ and $p$:

\begin{equation}
\label{G-Def}
G=\left\langle F_{N_1p_1}\ F_{N_2p_2} \right\rangle.
\end{equation}
The correlator $G$ is also multiplicatively renormalizable and,
as a consequence, it satisfies the differential RG equation
(\ref{IR-Scaling})--(\ref{IR-Scaling-Coeff}), which describes
the IR scaling behavior. But, due to the mixing condition of the
operators $F_{Np}$ themselves, the solution of this equation for
the function $G$ is more involved.

Since the correlator $G$ is a function of ${\bf x=r_1-r_2}$, $m$, $M$ and
$f$, the dimensionality representation for it is
\begin{equation}
\label{Corr-Dim}
G\propto
\nu^{d^\omega_G}\mu^{d_G}
\cdot \widehat{\Phi}\left(\mu r,\,mr,\,Mr,\,f\right),
\end{equation}
where $\mu$ is the renormalization mass, $\nu$ is the viscosity coefficient and $\widehat{\Phi}\left(\mu r,\,mr,\,Mr,\,f\right)$ is some function of four dimensionless parameters.
The differential operator ${\cal D}_{RG}$ in this case
reduces to the form

\begin{equation}
\label{D-RG-Correlator}
{\cal D}_{RG} =-{\cal D}_{\bf r}+ {\cal D}_{m} + {\cal D}_{M} + \gamma_f^*{\cal D}_{f}.
\end{equation}
Applying it to the correlator $G$ and denoting $F_{N_1p_1}$ as $F_i$
and $F_{N_2p_2}$ as $F_k$ (we recall, that $N_1$ may not be equal
to $N_2$, i.e., operators $F_i$ and $F_k$ may belong to different
renormalization sets), we obtain the differential equation

\begin{equation}
\label{G-RG-Eq}
{\cal D}_{RG}\ G_{ik}= \Delta_{is}G_{sk} + \Delta_{ks}G_{is},
\end{equation}
where $G_{ij}=\left\langle F_iF_j \right\rangle$, $\Delta_{ij}$
is the critical dimension of the correlator $G_{ij}$, and the summation
over repeated indices is implied. Note that due to the difference of
the numbers $N_1$ and $N_2$ in the initial operators $F_{Np}$ in
(\ref{G-Def}), the matrices $\Delta_{is}$ and $\Delta_{ks}$ in
(\ref{G-RG-Eq}) can have different dimensions.


Let us now consider the operators $\widetilde{G}_{ik}$ instead of $G_{ik}$,
namely, the correlation functions of operators $\widetilde{F}$
(see~(\ref{FF-tld})) that possess definite critical dimensions:

\begin{equation}
\label{G-Tilde-Def}
\widetilde{G}_{ik}=\left\langle \widetilde{F}_{i}\ \widetilde{F}_{k} \right\rangle.
\end{equation}
A few remarks follow about numbering and indices $i$ and $k$ in the
definition~(\ref{G-Tilde-Def}):

(1) The initial operator $F$ is defined in~(\ref{F-N-p}), namely

\begin{equation}
F_{Np}=(\theta_a\theta_a)^p\ (n_s\theta_s)^{2m}, \quad \ N=2(p+m).
\end{equation}

(2) Since at renormalization the operators, which can mix together,
have the same number $N$ (see~(\ref{F-diag1})), therefore for fixed
$N$ we may define a vector ${\bf F}$~(\ref{F-Vector-Def}), namely

\begin{equation}
\label{F-Vector-Def-Numb}
{\bf F}=
\begin{pmatrix}
F_1\\
F_2\\
\vdots \\
F_{N/2+1}
\end{pmatrix}
=
\begin{pmatrix}
(\theta_a\theta_a)^N\\
(\theta_a\theta_a)^{N-2}\cdot(n_s\theta_s)^2\\
\vdots \\
(n_s\theta_s)^N
\end{pmatrix}.
\end{equation}

(3) Let us define the vector ${\bf \widetilde{F}}$ as in~(\ref{FF-tld}),
namely

\begin{equation}
\label{FF-tld-App}
F^R_l=U_{lp}\widetilde{F}^R_p,
\end{equation}
where the matrix $U_{lp}$ is such that the matrix of critical dimensions
$ \widetilde{\Delta}_{F}= U_{F}^{-1} \Delta_{F} U_{F}$ is a Jordan matrix
(see sec.~\ref{sec:Diag}) and has the form~(\ref{U-UnDef}).

Therefore the operator $\widetilde{F}_i$ in the definition of the
correlation function~(\ref{G-Tilde-Def}) is not arbitrary, but is
constructed using~(\ref{FF-tld-App}) as a linear combination of the
operators $F_i$, whose numbering is strictly defined in
(\ref{F-Vector-Def-Numb}).

The correlation function $\widetilde{G}_{ik}$ satisfies the differential
equation in the form~(\ref{G-RG-Eq}), but with Jordan matrices
$\widetilde{\Delta}_{ik}$:

\begin{equation}
\label{G-RG-Eq-App-tilde}
{\cal D}_{RG}\ \widetilde{G}^R_{ik}= \widetilde{\Delta}_{is}
\widetilde{G}^R_{sk} + \widetilde{\Delta}_{ks}\widetilde{G}^R_{is}.
\end{equation}
If the operator $\widetilde{F}_i$, entering into the correlator
$\widetilde{G}_{ik}$, belongs to the set with number $N_1$, and the operator
$\widetilde{F}_k$ belongs to the set with number $N_2$, the expression
(\ref{G-RG-Eq-App-tilde}) is in fact a system of $(N_1/2+1)\times(N_2/2+1)$
nonseparable (due to nondiagonal, but Jordan form of matrices
$\widetilde{\Delta}_{ik}$) differential equations.

The matrices $\widetilde{\Delta}_{is}$ and $\widetilde{\Delta}_{ks}$ in
(\ref{G-RG-Eq-App-tilde}) have the form

\begin{equation}
\label{Delta-F-Tilde-App}
\widetilde{\Delta}_{F}
=
\begin{pmatrix}
\lambda_{1(2)} & 1 & 0 & \dots & 0 \\
0 & \lambda_{1(2)} & 1 &  & \vdots \\
\vdots & 0 & \ddots & \ddots  &0 \\
\vdots &  &  & \ddots & 1 \\
0 & \dots &  & 0 & \lambda_{1(2)}
\end{pmatrix},
\end{equation}
where $\lambda_1=-N_1$ and $\lambda_2=-N_2$ (see Appendix~\ref{App:I}).

Taking into account the expression~(\ref{Delta-F-Tilde-App}) it is obvious,
that if the both operators $\widetilde{F}_i$ and $\widetilde{F}_k$ are
not ``the last from the end,'' i.e., if $i\neq N_1/2+1$ and
$k\neq N_2/2+1$, then each of the terms in~(\ref{G-RG-Eq-App-tilde}) has
two contributions~-- one is the function $\widetilde{G}^R_{ik}$ with
coefficient $\lambda_{1(2)}$ and the other is either the function
$\widetilde{G}^R_{i+1,k}$ for the first term or the function
$\widetilde{G}^R_{i,k+1}$ for the second term,
both having coefficients~$1$. If one of the operators
$\widetilde{F}_i$ and $\widetilde{F}_k$ is ``the last from the end,''
i.e., if $i$ or $k$ is equal to $N_{1(2)}/2+1$, then this contribution
will be reduced to the only term $\widetilde{G}^R_{ik}$ with
the coefficient~$\lambda_{1(2)}$.

As a consequence, there is only one differential equation with one term
in the RHS, namely

\begin{equation}
\label{G-Tilde-Eq-Late}
{\cal D}_{RG}\ \widetilde{G}^R_{N_{1}/2+1\ N_{2}/2+1}=
(\lambda_1+\lambda_2)\cdot\widetilde{G}^R_{N_{1}/2+1\ N_{2}/2+1}.
\end{equation}
The solution of this RG equation is found in a standard way and has the following form:

\begin{equation}
\label{G-Tilde-0}
\widetilde{G}_0^R\equiv\widetilde{G}^R_{N_{1}/2+1\ N_{2}/2+1}
\propto (\mu r)^{-\lambda_1-\lambda_2}\cdot\Phi\left(1,\,mr,\,Mr,
\,\bar{f}\right),
\end{equation}
where 
$\bar{f}$ is the invariant charge 
and $\bar{f}\to fr^\xi$ as $1/\mu r\to0$, see~\cite{Vasiliev-Green}.

Then, if $i=N_1/2+1$ and $k=N_2/2$ or if $i=N_1/2$ and $k=N_2/2+1$,
i.e., if $k+i=(N_1+N_2)/2+1$, we have two equations of type
\begin{equation}
{\cal D}_{RG}\ \widetilde{G}^R_{1}= (\lambda_1+\lambda_2)
\cdot\widetilde{G}^R_{1}+\widetilde{G}_0^R,
\end{equation}
which involves the already known function $\widetilde{G}_0^R$ in the RHS.
Its solution contains a power factor and a polynomial of a logarithm,
i.e., up to a dimensional factor it is
\begin{equation}
\label{G-Tilde-1}
\widetilde{G}_1^R \propto (\mu r)^{-\lambda_1-\lambda_2}
\cdot P_{1}\left[\ln(\mu r)\right]\cdot\Phi\left(1,\,Mr,\,mr,
\,fr^\xi\right),
\end{equation}
where $P_{1}\left[\ln\mu r\right]$ is a first-degree polynomial of
the argument $\ln(\mu r)$. Using~(\ref{G-Tilde-0}) and~(\ref{G-Tilde-1})
we may write, that the asymptotic behavior of the sum
$\widetilde{G}_0^R+\widetilde{G}_1^R$ is the same as that of the function
$\widetilde{G}_1^R$ itself:

\begin{equation}
\widetilde{G}_0^R+\widetilde{G}_1^R\cong\widetilde{G}_1^R \propto (\mu r)^{-\lambda_1-\lambda_2}\cdot P_{1}\left[\ln\mu r\right]\cdot\Phi\left(1,\,Mr,\,mr, \,fr^\xi\right).
\end{equation}
Then,
if $k+i=(N_1+N_2)/2$, we have three expressions, which in the RHS involve
the function $\widetilde{G}_1$ that is already known from expression
(\ref{G-Tilde-1}), and may also involve the function $\widetilde{G}_0$,
that is also known:
\begin{equation}
{\cal D}_{RG}\ \widetilde{G}^R_{2}=
(\lambda_1+\lambda_2)\cdot\widetilde{G}^R_{2}+\widetilde{G}_1^R.
\end{equation}
Its solution contains a second-degree polynomial with the argument
$\ln(\mu r)$, i.e.,
\begin{equation}
\label{G-Tilde-2}
\widetilde{G}_2^R \propto (\mu r)^{-\lambda_1-\lambda_2}
\cdot P_{2}\left[\ln(\mu r)\right]\cdot\Phi\left(1,\,Mr,\,mr,
\,fr^\xi\right).
\end{equation}

The procedure is similar for the next functions.
It is obvious that the number of equations, which contain in the RHS a
function that is known from the previous step, increases for
$(N_1+N_2)/2+2\leq i+k\leq (N_1+N_2)/4+1$ and decreases
if $(N_1+N_2)/4+1\leq i+k\leq 2$. As a consequence, in this system
there is only one function, namely that with $i+k=2$, whose asymptotic
behavior contains a polynomial of the maximal power of the logarithm:
\begin{equation}
\label{G-Tilde-Max}
\widetilde{G}_{11}^R \propto (\mu r)^{-\lambda_1-\lambda_2}
\cdot P_{(N_1+N_2)/2}\left[\ln\mu r\right]
\cdot\Phi\left(1,\,Mr,\,mr, \,fr^\xi\right),
\end{equation}
where $P_{(N_1+N_2)/2}\left[\ln\mu r\right]$ is an $(N_1+N_2)/2$-degree
polynomial with the argument $\ln(\mu r)$.

Finally, expressions like~(\ref{G-Tilde-0}),~(\ref{G-Tilde-1}) and
(\ref{G-Tilde-Max}) give the asymptotic behavior of {\it any} function
$\widetilde{G}_{ik}^R$.

In order to obtain the asymptotic behavior of the correlation functions
of the initial operators ``without tilde,''
we have to use the expression~(\ref{FF-tld-App}). The inverse
matrix $U^{-1}$ has the form

\begin{equation}
\label{U-Inversed-App2}
U^{-1}_{F}
=
\begin{pmatrix}
0  & \dots & \dots & \dots & 0 & \hat{u}_{1n}\\
\vdots &  & & & \hat{u}_{2\ n-1} &  \hat{u}_{2n}\\
\vdots &  &  &~\reflectbox{$\ddots$} & \vdots & \vdots\\
\vdots & &~\reflectbox{$\ddots$} & \ & \vdots & \hat{u}_{n-2\ n}\\
0 & \hat{u}_{n-1\ 2} & \dots & \dots & \hat{u}_{n-1\ n-1} & \hat{u}_{n-1\ n}\\
\hat{u}_{n1} &  \dots & \dots & \hat{u}_{n,n-2} & \hat{u}_{n,n-1} & \hat{u}_{nn}
\end{pmatrix},
\end{equation}
wherein all the elements $\hat{u}_{ab}\neq0$. Note that the two operators,
entering in~(\ref{G-Tilde-Def}), bring about two (perhaps different)
matrices $U^{-1}_{F_i}$ and $U^{-1}_{F_k}$.


From the expression~(\ref{U-Inversed-App2}) it follows that the operators
$\widetilde{F}^R$ from the closed set $\left\{\widetilde{{\bf F}}^R\right\}$
with the dimension $N$ can be expressed in terms of operators $F^R$ of the
closed set $\left\{{\bf F}^R\right\}$ with the same dimension $N$ in the
following way:

\begin{equation}
\widetilde{F}_{1}^R \cong F_{N/2+1}^R
\end{equation}
(up to a numerical coefficient, namely $\hat{u}_{1n}$);

\begin{equation}
\widetilde{F}_{2}^R \cong F_{N/2}^R + F_{N/2+1}^R
\end{equation}
and so on, i.e., for {\it any} $i$

\begin{equation}
\widetilde{F}_{i}^R \cong \sum_{\alpha}F_\alpha^R + F_{N/2+1}^R,
\end{equation}
where $\alpha\neq N/2+1$ and numbers all other operators.

Now we are ready to find the desired asymptotic form of the correlation
function $G_{ik}$. Let us denote the elements of the matrix $U^{-1}_{F_i}$
for the operator $F_i$ in the correlator
$\widetilde{G}_{ik}=\left\langle \widetilde{F}_{i}\
\widetilde{F}_{k} \right\rangle$ as $\hat{u}_{ab}$,
the elements of matrix $U^{-1}_{F_k}$ for the operator $F_k$
as $\breve{u}_{ab}$, so that

\begin{widetext}
\begin{equation}
\label{G-Tilde-G-11}
\widetilde{G}_{11}^R = \hat{u}_{1,N_1/2+1}\breve{u}_{1,N_2/2+1}
\cdot G_{N_1/1+1\ N_2/1+1}^R;
\end{equation}

\begin{equation}
\label{G-Tilde-G-12}
\widetilde{G}_{12}^R = \hat{u}_{1,N_1/2+1}\cdot\left(\breve{u}_{2,N_2/2}
\cdot G_{N_1/1+1\ N_2/1}^R + \breve{u}_{2,N_2/2+1}\cdot
G_{N_1/1+1\ N_2/1+1}^R\right);
\end{equation}

\begin{equation}
\label{G-Tilde-G-13}
\widetilde{G}_{13}^R = \hat{u}_{1,N_1/2+1}\cdot\left(\breve{u}_{3,N_2/2-1}
\cdot G_{N_1/1+1,N_2/1-1}^R + \breve{u}_{3,N_2/2}\cdot G_{N_1/1+1,N_2/1}^R
+ \breve{u}_{3,N_2/2+1}\cdot G_{N_1/1+1,N_2/1+1}^R\right)
\end{equation}
\end{widetext}
and so on. Equations~(\ref{G-Tilde-G-11})--(\ref{G-Tilde-G-13}) show that the
expression for {\it any} function $\widetilde{G}_{ik}^R$ contains in the
RHS the function $G_{N_1/1+1\ N_2/1+1}^R$ with {\it different} coefficients
($\breve{u}_{a,b}\neq\breve{u}_{a+1,b}$ and
$\hat{u}_{a,b}\neq\hat{u}_{a+1,b}$ for all $a$, $b$), therefore the
expression for {\it any} function $G_{ik}^R$ contains in the RHS the
function $\widetilde{G}_{11}^R$.
This fact together with the expression~(\ref{G-Tilde-Max}) gives the
sought-for asymptotic behavior of the pair correlator function of the
initial operators from the set $\left\{{\bf F}\right\}$:

\begin{widetext}
\begin{equation}
G_{ik}^R\cong \widetilde{G}_{11}^R \propto
(\mu r)^{-\lambda_1-\lambda_2}\cdot P_{(N_1+N_2)/2}
\left[\ln\mu r\right]\cdot\Phi\left(1,\,Mr,\,mr, \,fr^\xi\right)
\ \ \ \ \forall i,k.
\end{equation}
\end{widetext}
Using the above written relations $\lambda_1=-N_1$ and $\lambda_2=-N_2$
we obtain the sought-for asymptotic behavior of the pair correlator
(\ref{G-Def}) up to a dimensional factor:
\begin{widetext}
\begin{equation}
\label{G-Asympt-General}
G_{ik}^R\cong \widetilde{G}_{11}^R \propto (\mu r)^{N_1+N_2}
\cdot P_{(N_1+N_2)/2}\left[\ln\mu r\right]\cdot\Phi
\left(1,\,Mr,\,mr, \,fr^\xi\right) \ \ \ \ \forall i,k.
\end{equation}
\end{widetext}
Here $P_L$ is a polynomial function of degree $L$ and $\Phi$ is a
function  of three dimensionless arguments. Its asymptotic behavior
is studied using the OPE.

\section{Operator product expansion and violation of scaling} \label{sec:OPE}

Representations~(\ref{G-Asympt-General}) for any scaling
functions $\breve{\Phi}\left(Mr,\ mr, \ fr^\xi\right)\equiv\Phi\left(1,\,Mr,\,mr, \,fr^\xi\right)$ describe the behavior
of the correlation functions
for $\mu r\gg1$ and any fixed value of $Mr$. The inertial range
$l\ll r\ll L$  corresponds to the additional condition $Mr\ll 1$.
The form of the functions $\Phi\left(Mr\right)$ is not determined by the
RG equations
themselves; in analogy with the theory of critical phenomena, its
behavior for $Mr\to0$ is studied using the well-known Wilson operator
product expansion (OPE).

According to the OPE, the equal-time product $F_{1}(x')F_{2}(x'')$
of two renormalized operators for
${\bf x}\equiv ({\bf x'} + {\bf x''} )/2 = {\rm const}$ and
${\bf r}\equiv {\bf x'} - {\bf x''}\to 0$ has the representation
\begin{equation}
\label{OPE}
F_{1}(x')F_{2}(x'')=\sum_{\widetilde{F}} C_{\tilde{F}} ({\bf r})
\widetilde{F}(t,{\bf x}) ,
\end{equation}
where the functions $C_{\tilde{F}}$ are coefficients regular in $M^{2}$
and $\widetilde{F}$ are all possible renormalized local composite operators
allowed by symmetry (more precisely, see below).
Without loss of generality, it can be assumed that the expansion
is made in the basis operators $\widetilde{F}$ of the type~(\ref{FF-tld}),
i.e., those
having definite critical dimensions $\widetilde{\Delta}_{F}$.
The renormalized correlator $\langle F_{1}(x)F_{2}(x') \rangle$
is obtained by averaging~(\ref{OPE}) with the weight
$\exp S_{R}$, where $S_{R}$ is the renormalized action~(\ref{RenAction}).
The quantities $\langle \widetilde{F}\rangle \propto
(Mr)^{\widetilde{\Delta}_{F}}$
appear on the right hand side.  Their asymptotic behavior
for $M\to0$ is found from the corresponding RG equations and
has the form

\begin{equation}
\langle \widetilde{F}_{\alpha}\rangle \propto
(Mr)^{\widetilde{\Delta}_{\alpha}},
\end{equation}
where $\widetilde{\Delta}_{F}$ is a Jordan matrix~(\ref{Delta-F-Tilde}) and
$(Mr)^{\widetilde{\Delta}_{\alpha}}$ is a matrix of type~(\ref{MR-Delta}).

Note that due to the form of the differential operator ${\cal D}_{RG}$
\eqref{D-RG-Correlator} the solution of the equation
\eqref{G-RG-Eq-App-tilde} implies the substitution $\mu r=1$, i.e.,
the matrix $(M/\mu)^{\widetilde{\Delta}_{\alpha}}$ given in~(\ref{MR-Delta}) is replaced by the matrix
$(Mr)^{\widetilde{\Delta}_{\alpha}}$.

From the operator product expansion~(\ref{OPE}) we therefore
find the following expression  for the scaling function
$\Phi\left(Mr,\ mr, \ fr^\xi\right)$ in the representation
(\ref{G-Asympt-General}) of the correlator
$\langle F_{1}(x)F_{2}(x') \rangle$:

\begin{equation}
\label{OPE-Asympt}
\bar{\Phi}\left(Mr\right)=\sum_{\alpha}A_{\alpha}\, (Mr)^{\Delta_{\alpha}},
\quad Mr \ll1,
\end{equation}
where the coefficients $A_{\alpha}=A_{\alpha}(Mr)$, coming from the Wilson
coefficients $C_{\alpha}$ in~(\ref{OPE}), are regular
in $(Mr)^{2}$. Here and below we do not distinguish the two
IR scales $M$ and $m$, first introduced in~\eqref{Cik} and
\eqref{VV-delta}, and
$\left.\bar{\Phi}\left(Mr\right)\equiv \breve{\Phi}\left(Mr,\
fr^\xi\right)\right|_{fr^\xi={\rm const}}$.

In general, the operators entering into the OPE are
those which appear in the corresponding Taylor expansions, and also
all possible operators that admix to them in renormalization
\cite{Zinn,Vasiliev-Green}. From~(\ref{F-Tilde-Asympt}) it is clear,
that the main contribution to the sum~(\ref{OPE-Asympt}) is given by the
operator $\widetilde{F}^R_1$, which possesses maximal singularity.
Therefore, combining the RG representation~(\ref{G-Asympt-General}) with
the OPE representation~(\ref{OPE-Asympt}) gives the desired asymptotic
expression for the pair correlation function $G$~(\ref{G-Def}) in the
inertial range:

\begin{widetext}
\begin{equation}
G=\left\langle F_{N_1\,p_1}\ F_{N_2\,p_2} \right\rangle \propto
(\mu r)^{N_1+N_2}\cdot (M r)^{-N_1-N_2}\cdot P_{(N_1+N_2)/2}\left[\ln \left(1/\mu r\right)\right]\cdot
P_{(N_1+N_2)/2}\left[\ln Mr\right]\cdot\widetilde{\Phi}\left(fr^\xi\right).
\label{G-tmp}
\end{equation}
Taking into account that canonical dimension $d_G=-N_1-N_2$, expression~\eqref{G-tmp} together with the dimensionality representation~\eqref{Corr-Dim} gives

\begin{equation}
G=\left\langle F_{N_1\,p_1}\ F_{N_2\,p_2} \right\rangle \propto
\nu^{d^\omega_G}\cdot M^{-N_1-N_2}\cdot P_{(N_1+N_2)/2}\left[\ln \mu r\right]\cdot
P_{(N_1+N_2)/2}\left[\ln Mr\right]\cdot\widetilde{\Phi}\left(fr^\xi\right),
\label{Vasik}
\end{equation}
where the leading term is

\begin{equation}
\label{Answer}
G \propto \nu   ^{d^\omega_G}\cdot M^{-N_1-N_2}\cdot \left[\ln \mu r\right]^{(N_1+N_2)/2}
\cdot \left[\ln Mr\right]^{(N_1+N_2)/2}\cdot\widetilde{\Phi}
\left(fr^\xi\right)
\end{equation}
\end{widetext}
with a certain scaling function $\widetilde{\Phi}\left(fr^\xi\right)$, restricted in the inertial range
$l\ll r\ll L$.

\section{Conclusion} \label{sec:Conc}

We applied the field theoretic renormalization group and the operator
product expansion to the analysis of the inertial-range asymptotic behavior
of a divergence-free vector field, passively advected by strongly
anisotropic random flow. The advecting velocity field was taken Gaussian,
not correlated
in time, with the given pair correlation function described by the
expressions~(\ref{ViVk})--(\ref{D-v}).
This ensemble can be viewed as the $d$-dimensional generalization
of the ensemble introduced in~\cite{AM} in the context of passive scalar
problem. Following~\cite{amodel}, we included into
the stochastic advection-diffusion equation~(\ref{stoch}) an additional
arbitrary parameter ${\cal A}$, so that the resulting model involves,
as special cases, the kinematic dynamo model for
magnetohydrodynamic turbulence, the linearized Navier--Stokes equation
and the case of passive vector ``impurity.''

In contrast to the famous Kraichnan's rapid-change model, where the
correlation functions exhibit anomalous scaling behavior with infinite sets
of anomalous exponents, here the dependence on the integral turbulence
scale $L$ demonstrates a logarithmic character:
the anomalies manifest themselves as polynomials of logarithms of $(L/r)$,
where $r$ is the separation.
The inertial-range asymptotic expressions for various correlation functions
are summarized in expressions~(\ref{Vasik}) and~(\ref{Answer}).

The key point is that the matrices of scaling dimensions of the relevant
families of composite fields (operators) appear nilpotent and cannot be
diagonalized and can only be brought to Jordan form; hence the logarithms.
The detailed technical proof of this fact is given. However,
we cannot give yet a clear physical interpretation of a logarithmic
violation of scaling behavior.

The possibility of logarithmic dependence of various correlation functions
on the integral scale $L$ and the separation $r$
should be taken into account in analysis of experimental data. Of course,
it is desirable to analyze the inertial-range behavior of more realistic
models, in particular, to introduce finite correlation time to the
correlation function of the velocity field. This work is in progress.

\section*{Acknowledgments}

The authors are indebted to L.~Ts.~Adzhemyan, Michal Hnatich, Juha Honkonen,
Arpine Kozmanyan, L.~N.~Lipatov, M.~Yu.~Nalimov, S.~L.~Ogarkov and
S.~A.~Paston for discussions.

The work was supported by the Saint Petersburg State University within the
research grant 11.38.185.2014 and by the Russian Foundation for Basic
Research within the project 12-02-00874-a.
N.M.G. was also supported by the D.~B.~Zimin's ``Dynasty'' foundation.

\appendix

\section{On the possibility of two different spatial scales in anisotropic
vector models}
\label{App:III}

Consider the action functional~\eqref{Action} and recall the transversality
conditions~(\ref{transvers}). Those conditions introduce some important
difference between the present model and
its scalar analog, studied in~\cite{AntMal2011} within the RG+OPE approach.
Let us decompose the vector field $\boldsymbol{\theta}$ into the
components parallel and perpendicular to the vector ${\bf n}$:
$\boldsymbol{\theta}={\bf n}\varphi+\boldsymbol{\theta_\perp}$,
where $(\boldsymbol{\theta_\perp}\cdot{\bf n})=0$, and similarly for
$\boldsymbol{\theta}'$.

Now let us try to define, in analogy with the scalar case, one temporal
scale and two independent spatial scales that correspond to the directions
parallel and perpendicular to ${\bf n}$:
\begin{equation}
\label{Kuriza}
[F] \sim [T]^{-d_{F}^{\omega}}  [L_{\bot}]^{-d_{F}^{\bot}}
[L_{\parallel}]^{-d_{F}^{\parallel}}.
\end{equation}
The normalization conditions following from the definition (\ref{Kuriza})
have the forms:
\begin{subequations}
 \label{TwoDim}
 \begin{equation}
   d_{k_{\bot}}^{\bot}= -d_{\bf x_{\bot}}^{\bot}=1,\ \ \
d_{k_{\bot}}^{\parallel}=-d_{\bf x_{\bot}}^{\parallel}=0,
 \end{equation}
 \begin{equation}
  d_{k_{\parallel}}^{\parallel}= -d_{\bf x_{\parallel}}^{\parallel}=1,
\ \ \ d_{k_{\parallel}}^{\bot}=-d_{\bf x_{\parallel}}^{\bot}=0.
 \end{equation}
\end{subequations}

The other dimensions are determined by the requirement that all terms in the
action functional be dimensionless (with respect to all three independent
dimensions). In particular, this requirement for the term
\begin{equation}
\label{B19}
\theta'_i\partial_t\theta_i=\int dt\int d{\bf x}\
\theta'_i\partial_t\theta_i
\end{equation}
gives (from now on we discuss only spatial dimensions):
\begin{subequations}
 \label{DimThetaThetaPrime}
  \begin{equation}
  \label{15a}
   -1+d^\parallel_{\varphi}+d^\parallel_{\varphi'}=0,
 \end{equation}
  \begin{equation}
  \label{15b}
   -1+d^\parallel_{\theta_\perp}+d^\parallel_{\theta_\perp'}=0,
 \end{equation}
  \begin{equation}
  \label{15c}
   1-d+d^\perp_{\varphi}+d^\perp_{\varphi'}=0,
 \end{equation}
 \begin{equation}
 \label{15d}
   1-d+d^\perp_{\theta_\perp}+d^\perp_{\theta_\perp'}=0.
 \end{equation}
 \end{subequations}

The full set of such equations for all terms in the action functional
has the unique solution which coincides
exactly with the case of scalar model~\cite{AntMal2011}.
In particular, the dimensions of the fields $\varphi$ and
$\boldsymbol{\theta}_\perp$ are identical and coincide with their
analogs of the scalar field, and similarly for the fields $\varphi'$ and
$\boldsymbol{\theta}'_\perp$ (note that, e.g., the equations
(\ref{15a}) and (\ref{15c}) coincide with the equations
(\ref{15b}) and (\ref{15d}), respectively).

However, those dimensions do not satisfy the additional restrictions,
imposed by the transversality conditions~\eqref{transvers}:
\begin{subequations}
  \begin{equation}
  \label{Transverse-Perp}
   d^\perp_{\varphi}=1+d^\perp_{\theta_\perp},
 \end{equation}
  \begin{equation}
  \label{Transverse-Par}
   1+d^\parallel_\varphi=d^\parallel_{\theta_\perp},
 \end{equation}
  \begin{equation}
   \label{Transverse-Prime-Perp}
   d^\perp_{\varphi'}=1+d^\perp_{\theta'_\perp},
 \end{equation}
  \begin{equation}
    \label{Transverse-Prime-Par}
   1+d^\parallel_{\varphi'}=d^\parallel_{\theta'_\perp}.
 \end{equation}
\end{subequations}
Indeed, taking into account expressions~(\ref{15a})--(\ref{15d}), from
(\ref{Transverse-Perp}) and~(\ref{Transverse-Par}) it follows that
\begin{subequations}
   \begin{equation}
    d^\perp_{\varphi'}=-1+d^\perp_{\theta'_\perp},
 \end{equation}
 \begin{equation}
    1+d^\parallel_{\varphi'}=2+d^\parallel_{\theta'_\perp},
 \end{equation}
\end{subequations}
which contradicts the relations~(\ref{Transverse-Prime-Perp})
and~(\ref{Transverse-Prime-Par}).

We conclude that in the present vector model independent dimensions for the
transverse and longitudinal directions cannot be introduced, and only the
total canonical dimensions
 \begin{equation}
  d_F=d_F^k+2d_F^\omega,\quad  d_F^k=d_F^\bot+d_F^\parallel
 \end{equation}
make sense. As a consequence, the constant $f_0$, introduced to split $O_d$
symmetry of the Laplace operator
($\partial^2=\boldsymbol{\partial}^2_\perp+f_0\cdot\partial^2_\parallel$),
in our model appears dimensionless in contrast to~\cite{AntMal2011}.

However, independent spatial dimensions can indeed be introduced in the
modified vector model, in which the transversality
conditions~(\ref{transvers})
are satisfied due to simultaneous vanishing of their both terms,
that is
\begin{equation}
\partial_{\parallel}\varphi = - \partial_{i}^{\bot}\theta_{i}^{\bot}=0,
\end{equation}
and similarly for $\boldsymbol{\theta}'$. In other words, the field
$\varphi=\varphi(t,{\bf x}_\perp)$ is independent of the
longitudinal coordinate $x_{\parallel}$, while $\boldsymbol{\theta_\perp}$
is ortogonal to ${\bf n}$ and to the momentum ${\bf k}$. Then the conditions
(\ref{Transverse-Perp})--(\ref{Transverse-Prime-Par}) no longer follow
from~(\ref{transvers}). It would be interesting to study such model,
because its asymptotic behavior can be essentially different from that
of the present case.

\section{Derivation of the propagator. A simple model}
\label{Sec:Toymodel}

In order to justify our scheme of derivation of the
propagator~\eqref{Propgtr-Full-WithPart}
in sec.~\ref{sec:TruePropagator}, consider a
simpler example, a toy ``field theory'' of a single constant (i.e.,
independent on $x$) real random $n$-component vector
{\mbox{\boldmath $\theta$}}$=\{\theta_{1}, \dots, \theta_{n}\}$ with the
action function
\begin{equation}
{\cal S}(\theta) = - \frac{1}{2} \theta_{i} M_{ij} \theta_{j},
\label{toi1}
\end{equation}
where $M_{ij}$ is a positive real symmetric matrix (so that $\det M>0$),
and the summations over the vector indices from 1 to $n$ are implied.
The generating function of the ``correlation functions''
$D_{ij}=\langle\theta_{i}\theta_{j}\rangle$ is
\begin{equation}
G(A) = C\cdot \prod_{i=1}^{n} \int_{-\infty}^{\infty} d\theta_{i}
\exp \{ {\cal S}(\theta) + \theta_{i}A_{i} \},
\label{toi2}
\end{equation}
where $A_{i}$ is the ``source'' and the normalization constant $C$ is chosen
such that $G(0)=1$. The expression~\eqref{toi2} is a Gaussian integral, so
\begin{equation}
G(A) =\exp \left\{ \frac{1}{2}A_{i}M^{-1}_{ij}A_{j} \right\}.
\label{toi2a}
\end{equation}
Now let us assume that the orthogonality condition
$\kappa_{i}\theta_{i}=0$ is imposed on the random variable with a certain
constant unit vector {\mbox{\boldmath $\kappa$}}$=\{\kappa_{i}\}$
(this is an analog of the transversality conditions~(\ref{Trans-Def})).
Then the generating functional should be understood as
\begin{equation}
G(A) = C\cdot \prod_{i=1}^{n}
\int_{-\infty}^{\infty} d\theta_{i}
\delta(\kappa_{i}\theta_{i})
\exp \{{\cal S}(\theta) + \theta_{i}A_{i} \},
\label{toi3}
\end{equation}
which is an analogue of the expression~\eqref{Delta-Green} for complete model~\eqref{Action}.

In order to calculate the integral in (\ref{toi3}), it is convenient to
choose the coordinate system such that the vector {\mbox{\boldmath $\kappa$}}
be oriented along the first axis,
{\mbox{\boldmath $\kappa$}} $=\{1,0,\dots,0\}$. The integration over
$\theta_{1}$ is readily performed owing to the factor $\delta(\theta_{1})$,
and the expression (\ref{toi3}) becomes
\begin{widetext}
\begin{equation}
G(A) = C\cdot \prod_{i=2}^{n} \int_{-\infty}^{\infty} d\theta_{i}
\exp \left\{ - \frac{1}{2} \theta_{i} \widetilde{M}_{ij} \theta_{j}
+ \theta_{i}A_{i} \right\} =
\exp \left\{ \frac{1}{2}A_{i}\widetilde{M}^{-1}_{ij}A_{j} \right\},
\label{toi4}
\end{equation}
\end{widetext}
where all the summations run from 2 to $n$ and $\widetilde{M}_{ij}$ is the
$(n-1)\times(n-1)$ matrix obtained from $M_{ij}$ by removing the uppermost
row and the leftmost column. The last equality is the direct analog
of (\ref{toi2a}) for the $(n-1)$-component field
$\{\theta_{2}, \dots, \theta_{n}\}$, and $\widetilde{M}^{-1}_{ij}$ is its
propagator: $D_{ij}=\widetilde{M}^{-1}_{ij}$ for $i,j=2,\dots,n$.
The propagators involving the component $\theta_{1}$
vanish: $\langle\theta_{1}\theta_{i}\rangle=0$ for all $i$.

In a covariant way (not related the special choice of the coordinates)
this procedure can be described in terms of the transverse projector
$P_{ij}=\delta_{ij}-\kappa_{i} \kappa_{j}$ (with respect to the vector
{\mbox{\boldmath $\kappa$}}).
Note that for our special choice of the coordinates it takes the form of
a diagonal matrix with the elements $\{0,1\dots,1\}$. Then the propagator
matrix $D_{ij}$ of the full $n$-component field, derived above, can be
obtained as follows. Consider the $n\times n$ matrix
$\widehat{M}_{ij} = P_{ik}M_{ks}P_{sj}$; its elements coincide with those of
the matrix $\widetilde{M}_{ij}$ for $i,j \ge2$ and vanish otherwise. Clearly,
this matrix cannot be inverted in the full $n$-dimensional space, but it
can be inverted on the subspace orthogonal to {\mbox{\boldmath $\kappa$}}.
In other words, the propagator matrix $D_{ij}$ is obtained from the
relation
\begin{equation}
\widehat{M}_{il} D_{lj} = P_{ik}M_{ks}P_{sl} D_{lj} = P_{ij},
\label{toi5}
\end{equation}
since the projector $P_{ij}$ on that subspace acts as the identity matrix.

It is probably worth to discuss alternative way of derivation the propagator
matrix: one can represent the $\delta$-function in (\ref{toi4}) by the
Fourier transform, which introduces additional integration variable $\wp$:
\begin{widetext}
\begin{equation}
G(A) = C\cdot \prod_{i=1}^{n} \int_{-\infty}^{\infty} d\theta_{i}
\int d\wp\, \exp \left\{ - \frac{1}{2} \theta_{i} M_{ij} \theta_{j}
-{\rm i}\wp(\kappa_{i}\theta_{i}) + \theta_{i}A_{i} \right\}.
\label{toi6}
\end{equation}
\end{widetext}
Now the propagators of the extended set of ``fields''
{\mbox{\boldmath $\theta$}}, $\wp$ are found in a standard fashion,
by inverting the (symmetrized) matrix entering the full action,
${\cal S(\theta)}-{\rm i}\wp(\kappa_{i}\theta_{i})$. Thus one has
to solve the equation
\begin{equation}
\label{toi7}
\begin{pmatrix}
M_{ij} & {\rm i}\kappa_{s} \\
{\rm i}\kappa_{l}  & 0 \\
\end{pmatrix}
\times
\begin{pmatrix}
D_{ij} & a_{s} \\
a_{l}  & b \\
\end{pmatrix}
=
\begin{pmatrix}
\delta_{ij} & 0 \\
0  & 1 \\
\end{pmatrix} ,
\end{equation}
where $D_{ij}$ is the sought-for propagator matrix for the $n$-component
field {\mbox{\boldmath $\theta$}}, $a_{i}=\langle \wp\theta_{i}\rangle$ and
$b=\langle \wp\wp \rangle$. In the component notation (\ref{toi7}) gives:
\begin{subequations}
\label{toi8}
\begin{align}
\label{toi8a}
M_{ik}D_{kj}+{\rm i}\kappa_{i}a_{j}=\delta_{ij},\\
\label{toi8b}
\kappa_{s}D_{si}=0,\\
M_{ik}a_{k}+{\rm i}\kappa_{i}b=0, \\
\kappa_{i}a_{i} =1.
\end{align}
\end{subequations}

From the equation~\eqref{toi8b} one can see that the propagator matrix is
transverse, i.e., $D_{kj}=P_{ks}D_{kj}$. Substituting this relation
into the first equation~(\ref{toi8a}) and multiplying it from the left
by $P_{li}$ gives the relation (\ref{toi5}). The remaining two equations
determine the propagators with the auxiliary field $\wp$.

To avoid possible confusion, we emphasize that the propagator matrix $D_{ij}$
obtained from the above procedure and satisfying relation (\ref{toi5})
differs, {\it in general} (e.g., for the anisotropic model~\eqref{V-n}, \eqref{Action}), from the expression $P_{ik}M^{-1}_{ks}P_{sj}$
that would be obtained from the integral (\ref{toi2}) if the source was chosen
to satisfy the relation $\kappa_{i}A_{i}=0$.

\section{The nilpotency of the anomalous dimension matrix}
\label{App:I}
\subsection{Definitions and aims}

In this section we will prove the nilpotency of the matrix $\gamma_F^*$
from~(\ref{gamma-F-fixed}) and, as a consequence, the Jordan form of the
critical dimension matrix $\Delta_{Np,\,Np'}$ from~(\ref{Crit-Dim-F}). Let
us recall some definitions and facts from sec.~\ref{sec:AmonDim}
and sec.~\ref{sec:Diag}.

Let us define the vector ${\bf F}$ as in~(\ref{F-Vector-Def}), namely

\begin{equation}
\label{F-Vector-Def-App}
{\bf F}=
\begin{pmatrix}
(\theta_i\theta_i)^N\\
(\theta_i\theta_i)^{N-2}\cdot(n_s\theta_s)^2\\
\vdots \\
(n_s\theta_s)^N
\end{pmatrix};
\end{equation}
the relation $F_i=Z_{ik}F_k^R$ between the set of unrenormalized operators
$\left\{F\right\}$ and the set of renormalized operators $\left\{F^R\right\}$
takes the form

\begin{widetext}
\begin{equation}
\label{F-Z-Fr-App}
\begin{pmatrix}
(\theta_i\theta_i)^N\\
(\theta_i\theta_i)^{N-2}\cdot(n_s\theta_s)^2\\
(\theta_i\theta_i)^{N-4}\cdot(n_s\theta_s)^4\\
\vdots \\
\vdots \\
(\theta_i\theta_i)^{2}\cdot(n_s\theta_s)^{N-2}\\
(n_s\theta_s)^N
\end{pmatrix}
=
\begin{pmatrix}
a_{11} & a_{12} & a_{13} & 0 & \dots & 0\\
a_{21} & a_{22} & a_{23} & a_{24} & &  \vdots\\
0 & a_{32} & a_{33} & a_{34} & \ddots & 0\\
\vdots & 0 & a_{43} & \ddots & \ddots & a_{n-2n}\\
\vdots &  &  & \ddots & \ddots & a_{n-1n}\\
0 &\dots & \dots & 0 & a_{nn-1} & a_{nn}
\end{pmatrix}
\cdot
\begin{pmatrix}
\left\{(\theta_i\theta_i)^N\right\}^R\\
\left\{(\theta_i\theta_i)^{N-2}\cdot(n_s\theta_s)^2\right\}^R\\
\left\{(\theta_i\theta_i)^{N-4}\cdot(n_s\theta_s)^4\right\}^R\\
\vdots \\
\vdots \\
\left\{(\theta_i\theta_i)^{2}\cdot(n_s\theta_s)^{N-2}\right\}^R\\
\left\{(n_s\theta_s)^N\right\}^R
\end{pmatrix}.
\end{equation}
\end{widetext}
Note that in the matrix $\hat{Z}$ the power $p$ of the operator $F_{Np}$
decreases from the right to the left.

Let us define $y$ as

\begin{equation}
y=-\frac{{\cal A}^2\cdot f}{2(d-2+{\cal A})}\cdot\xi.
\end{equation}
According to~(\ref{gamma-F-fixed}), the elements of the matrix of anomalous
dimensions 
$\hat{\gamma}_F=\hat{Z}_{F}^{-1}{\cal D}_{\mu }\hat{Z}_{F}$ at the fixed point $g^*$ are
\begin{subequations}
\label{gamma-F-fixed-App}
\begin{align}
\label{gamma-p+1-App}
\gamma_{Np,\,Np'+1}^*=&\ 2m(2m-1)\cdot y;\\
\label{gamma-p-App}
\gamma_{Np,\,Np'}^*=&\ (2p+8pm-2m(2m-1))\cdot y;\\
\label{gamma-p-1-App}
\gamma_{Np,\,Np'-1}^*=&\ (4p(p-1)-2p-8pm)\cdot y;\\
\label{gamma-p-2-App}
\gamma_{Np,\,Np'-2}^*=&\ (-4p(p-1))\cdot y,
\end{align}
\end{subequations}
and the critical dimension matrix for the operators $F_{Np}$ has the form
\begin{equation}
\label{Crit-Dim-F-App}
\Delta_{Np,\,Np'}=-2(p+m)\cdot\delta_{pp'}+\hat{\gamma}^*_{Np,\,Np'}.
\end{equation}
Here $-2(p+m)$ is its canonical dimension, $\delta_{pp'}$ is Kronecker's
$\delta$-symbol and $\hat{\gamma}^*_{Np,\,Np'}$ is the value of anomalous dimension matrix
at the fixed point.

The aim is to prove the nilpotency of the matrix $\hat{\gamma}_F^*$ from
(\ref{gamma-F-fixed-App}) and the Jordan form of the matrix $\Delta_{Np,\,Np'}$
from~(\ref{Crit-Dim-F-App}).
We will present the explicit expression for the matrix $U_N$
that brings the
matrix $\Delta_{F}$ to the Jordan form $\widetilde{\Delta}_{F}$ by
the transformation

\begin{equation}
\Delta_{F}= U_{N} \widetilde{\Delta}_{F} U_{N}^{-1}.
\end{equation}
As the number $N$ in~(\ref{F-Vector-Def-App}) may be arbitrary,
the dimension of the matrix $\hat{Z}_F$ in equation~(\ref{F-Z-Fr-App}) and,
as a consequence, of the matrices $\hat{\gamma}_F$ and $U_N$, namely
$(N/2+1)\times (N/2+1)$, also may be arbitrary. This means, that the
expression~(\ref{gamma-F-fixed-App}) gives us the algorithm to
construct the matrix $\hat{\gamma}_F^*$ for the set of initial operators
$\left\{F\right\}$ with arbitrary $N$~-- simply it gives the value
of each matrix element. And the difficulty and the fascination of
this task is to find an algorithm for constructing the 
matrix $U_N$, which brings it to the Jordan form,
applicable to an arbitrary number $N$, or, equivalently,
to the matrix $\hat{\gamma}_F^*$ with arbitrary dimension. Note, that if the
matrix $\Delta_{F}$ was diagonalizable, the diagonalizing
matrix $U_N$ would be unique for each {\it fixed} number $N$, but since the
matrix $\widetilde{\Delta}_{F}$ has the Jordan form, the matrix
$U_N$, which brings it to the Jordan form, is not unique for any fixed number $N$. Therefore, we will show
one of the possible forms of the matrix $U_N$, which brings the
matrix $\Delta_{F}$ to Jordan form and thus solves our problem.

Since each element of the matrix $\hat{\gamma}_F^*$ is a multiple to the scalar
number $y$, the nilpotency of the matrix $\hat{\gamma}_F^*$ is equivalent to the
nilpotency of the matrix $\hat{\epsilon}_F^*$, where
$y\cdot\hat{\epsilon}_F^* = \hat{\gamma}_F^*$.

\subsection{Motivation and idea}

Let us write the $3\times3$ ($N=4$) matrix $\hat{\epsilon}_F^*$ denoted as $A_4$:
\begin{equation}
A_4=
\begin{pmatrix}
4 & 4 & 8\\
2 & 8 & -10\\
0 & 12 & -12
\end{pmatrix}.
\end{equation}
It is nilpotent, its eigenvalues are
\begin{equation}
\lambda_1=\lambda_2=\lambda_3=0.
\end{equation}
The matrix $U_4$, which brings the matrix $A_4$ to the Jordan form, is built from the
eigenvectors of the matrix $A_4$. Find them:
\begin{equation}
V_1=
\begin{pmatrix}
1\\
1\\
1
\end{pmatrix};
\quad
V_2=
\begin{pmatrix}
1/6\\
1/12\\
0
\end{pmatrix};
\quad
V_3=
\begin{pmatrix}
1/24\\
0\\
0
\end{pmatrix}.
\end{equation}
Note that the eigenvectors are determined by the condition
$(A_4-\lambda I)V_{i+1}=V_{i}$, which has a unique solution up to an
additional constant.

Thus the matrix $U_4$ takes the form
\begin{equation}
U_4=
\begin{pmatrix}
1 & 1/6 & 1/24\\
1 & 1/12 & 0\\
1 & 0 & 0
\end{pmatrix}
\quad \text{and}\quad
J_4=U_4^{-1}A_4U_4=
\begin{pmatrix}
0 & 1 & 0\\
0 & 0 & 1\\
0 & 0 & 0
\end{pmatrix}.
\end{equation}
Here one can notice an interesting property:
the product $A_4\cdot U_4$ is the same as the matrix $U_4$, but with all
columns shifted by one position to the right, namely
\begin{equation}
	A_4\cdot U_4=
	\begin{pmatrix}
	0 & 1 & 1/6\\
	0 & 1 & 1/12\\
	0 & 1 & 0
	\end{pmatrix}.
\end{equation}
Now, if we multiply the matrix $U_4^{-1}$ by the product $A_4\cdot U_4$,
it brings $A_4$ to the Jordan form:
\begin{equation}
U_4^{-1}\cdot
\begin{pmatrix}
1 & 1/6 & 1/24\\
1 & 1/12 & 0\\
1 & 0 & 0
\end{pmatrix}
=
\begin{pmatrix}
1 & 0 & 0\\
0 & 1 & 0\\
0 & 0 & 1
\end{pmatrix},
\end{equation}
but
\begin{equation}
\label{Jord-Inv-Matr}
U_4^{-1}\cdot
\begin{pmatrix}
0 & 1 & 1/6\\
0 & 1 & 1/12\\
0 & 1 & 0
\end{pmatrix}
=
\begin{pmatrix}
0 & 1 & 0\\
0 & 0 & 1\\
0 & 0 & 0
\end{pmatrix}.
\end{equation}
The feature~(\ref{Jord-Inv-Matr}) is not characteristic only for specific
matrices, but is a common rule. For any $M\times M$ nondegenerate matrix
\begin{equation}
\widehat{M}=
\begin{pmatrix}
a_{11} & a_{12} & \dots & a_{1n}\\
a_{21} & a_{22} & \dots & a_{2n}\\
\vdots &  & \ddots & \vdots\\
a_{n1} & a_{n2} & \dots & a_{nn}
\end{pmatrix}
\end{equation}
the product of $\widehat{M}^{-1}$ and $\widetilde{M}$, where $\widetilde{M}$
is the matrix $\widehat{M}$ with all columns shifted by one position to
the right and with all elements of the first column being equal to zero,
is a matrix of Jordan form:
\begin{equation}
\label{M-Shift-Common}
\widehat{M}^{-1}\cdot\widetilde{M}=\widehat{M}^{-1}\cdot
\begin{pmatrix}
0 & a_{11} & \dots & a_{1n-1}\\
0 & a_{21} & \dots & a_{2n-1}\\
\vdots &  & \ddots & \vdots\\
0 & a_{n1} & \dots & a_{nn-1}
\end{pmatrix}
=\begin{pmatrix}
0 & 1 & & \\
 & \ddots & \ddots & \\
&  & \ddots & 1\\
 & & & 0
\end{pmatrix}.
\end{equation}
Here empty space denotes the elements, which are equal to zero.
The expression
(\ref{M-Shift-Common}) is obvious. Multiplying $\widehat{M}^{-1}$ with the
first empty column gives us an empty column in the RHS. Multiplying
$\widehat{M}^{-1}$  with the other columns with numbers $2,\ \dots,\ n$
gives us the unity matrix, which however starts not from the cell $11$,
but from the cell $12$~-- i.e., the ``unity'' matrix with nonzero
terms not on the main diagonal, but on the diagonal above it.

Thus the idea is to find such a matrix $U_N$ with $\det U_N\neq0$, that
makes the product $A_N\cdot U_N$ equivalent to matrix $U_N$ itself, but
with its columns shifted as $i\to i+1$, and with the elements of the first
column equal to zero. If we find it, our problem will be solved:
\begin{equation}
U_N^{-1}\cdot\left[A_N\cdot U_N\right]=
\begin{pmatrix}
0 & 1 & & \\
 & \ddots & \ddots & \\
&  & \ddots & 1\\
 & & & 0
\end{pmatrix}.
\end{equation}

\subsection{Explicit form of the matrix $U_N$}

The next step is to understand the explicit form of the matrix $U_N$.
To this end, let us write the $10\times10$ matrix $\hat{\epsilon}_F^*$ for
$N=18$, denoted as $A_{18}$, and the 
matrix $U_{18}$, which brings it to the Jordan form (which
is found by direct calculation):

\begin{widetext}
\begin{equation}
\label{A-Expl}
A_{18}=
\begin{pmatrix}
18 & 270 & -288 & & & & & & & \\
2 &78 & 144 & -224 & & & & & & \\
 & 12 & 114& 42& -168& & & & & \\
 & &30&126 &-36 &-120 & & & & \\
 & & & 56&114 &-90 &-80 & & & \\
 & & & &90 &78 &-120 &-48 & & \\
 & & & & &132 &18 &-126 &-24 & \\
 & & & & & &182 &-66 &-108 &-8 \\
 & & & & & & & 240&-174 &-66 \\
 & & & & & & & & 306&-306 \\
\end{pmatrix},
\end{equation}

\begin{equation}
\label{U-Expl}
U_{18}=
\begin{pmatrix}
1 &9/306 & 36/73440 &84/{\rm III}\cdot182 &126/{\rm IV}\cdot132 &126/{\rm V}\cdot90 &84/{\rm VI}\cdot56 &36/{\rm VII}\cdot30 &9/{\rm VIII}\cdot12 & 1/{\rm IX}\cdot2 \\
1 &8/306 & 28/73440 &56/{\rm III}\cdot182 &70/ &56/ &28/ &8/ &1/ & \\
1 &7/306 & 21/73440 & 35/{\rm III}\cdot182& 35/& 21/& 7/& 1/& & \\
1 &6/306 & 15/73440 & 20/{\rm III}\cdot182& 15/&6/ & 1/& & & \\
1 &5/306 & 10/73440 &10/{\rm III}\cdot182 &5/ & 1/& & & & \\
1 &4/306 & 6/73440  & 4/{\rm III}\cdot182& 1/& & & & & \\
1 &3/306 & 3/73440  & 1/{\rm III}\cdot182& & & & & & \\
1 &2/306 & 1/73440  & & & & & & & \\
1 &1/306 &          & & & & & & & \\
1 &      &          & & & & & & & \\
\end{pmatrix}.
\end{equation}
\end{widetext}
Here the Roman figures denote the denominators from previous columns,
i.e., $\text{III}=73440$, $\text{IV}=73440\cdot182=13366080$, etc.
As all denominators in one column are identical, the symbol ``$/$''
denotes division of the numerator by the denominator, written in the
first element of the column.

From the explicit expression~(\ref{U-Expl}) it is obvious, that the
denominators of the elements of matrix $U_{18}$ are products of the
elements from the diagonal below the main diagonal of the matrix
$A_{18}$ from~(\ref{A-Expl}), and the numerators are the elements from
Pascal's triangle, namely $n\choose k$, where $n$ is the number of the
row (with numeration going from bottom up) and $k$ is the number of the
column (with numeration going from the left to the right).

Here ${n\choose k}\equiv C^k_n$ is the number of $k$ combinations
from the set of $n$ elements.

This is the conjecture, which we have to prove: the matrix, constructed
by the described rules is the sought-for matrix $U_N$ for {\it any} dimension
of initial matrix $A_N$ (i.e., for the family of operators with
{\it any} $N$).

One remark follows, which will be useful later: since in notation
(\ref{F-Z-Fr-App}) each row of matrix $A$ corresponds to an operator
with fixed number $p$, thus each element $n\choose k$ is actually
$p\choose {\cal C}$, where ${\cal C}$ is the number of the column
(starting from zero).

\subsection{The proof of our assumptions}

The proof is divided into several steps: first, we will prove the
reliability of the first two columns of the matrix, then the reliability
of the three lower diagonals. Finally, we will prove it for all the
other elements.

\subsubsection{The first column ({\cal C}=0)}
\label{sec:I-Col}

From expressions~(\ref{gamma-F-fixed-App}) it follows, that
$\sum_i\gamma^*_{Np,\,Np+i}=0$. This is the reason why in the case when the
first column of matrix $U_N$ is
$\begin{pmatrix}
1 \\
1 \\
\vdots\\
1
\end{pmatrix},$
the first column of matrix $A\cdot U$ is
$\begin{pmatrix}
0 \\
0 \\
\vdots\\
0
\end{pmatrix}$.

\subsubsection{The second column ({\cal C}=1)}

Now to have the base for further steps, we need to prove our conjecture for
the second column of matrix $U_N$, which is the first nontrivial column.

The latest element in the second column (in~(\ref{U-Expl}) it is $1/306$)
is the element, which is determined by the last element of the diagonal,
located below the main diagonal of matrix $A$. In~(\ref{A-Expl}) it is
equal to $-306$. Since this element is located on the aforementioned
diagonal, it is formed by the condition~(\ref{gamma-p+1-App}).
From the word ``the latest'' it follows, that this element corresponds
to the operator with $p=0$ and $2m=N$, therefore the
required element of the matrix $A_N$ is equal to $N(N-1)$
(for any dimension of matrix $A_N$). Therefore, the equation for the
element of the matrix $U_N$ (let us call it $X$, since it is what we
want to find) is

\begin{equation}
\label{I1-rule}
N(N-1)\cdot X=1.
\end{equation}
Therefore
\begin{equation}
\label{I1}
X=\frac{1}{N(N-1)},
\end{equation}
which is in agreement with~(\ref{U-Expl}), since from~(\ref{I1}) it follows
that for $N=18$ the element $X$ is equal to $1/306$.

An equation like~(\ref{I1-rule}), which describes the second element in the
second column, is

\begin{equation}
\label{I2-rule}
(N-2)(N-3)\cdot X+\frac{2+(N-2)(7-N)}{N(N-1)}=1.
\end{equation}
This follows from the requirement that the sum of the two terms
(corresponding to the transition with $\gamma^*_{Np,\,Np'+1}$
(\ref{gamma-p+1-App}) and $\gamma^*_{Np,\,Np'}$~(\ref{gamma-p-App}))
be equal to $1$, and from the observation, that these elements correspond
to the operator with $p=1$. From expression~(\ref{I2-rule}) it follows, that

\begin{equation}
\label{I2}
X=\frac{2}{N(N-1)}.
\end{equation}

The element, that is the third from the end, is governed by the sum of
three terms, constructed like expressions~(\ref{I1-rule}) and
(\ref{I2-rule}). As we go one position up, the parameters for the
operator $(\theta_i\theta_i)^p(n_s\theta_s)^{2m}$ become $p=2$ and $2m=N-4$.
So, 

$$(N-4)(N-5)\cdot X+$$
$$+\left[4+8(N-4)-(N-4)(N-5)\right]\frac{2}{N(N-1)}+$$
\begin{equation}
\label{I3-rule}
+\left[8-4-8(N-4)\right]\frac{1}{N(N-1)}=1,
\end{equation}
and hence

\begin{equation}
\label{I3}
X=\frac{3}{N(N-1)}.
\end{equation}

Expressions~(\ref{I1-rule}),~(\ref{I2-rule}) and~(\ref{I3-rule}) are
constructed from different number of terms, therefore they need to be
considered separately. Another distinguished element is the first element,
$9/306$ in expression~(\ref{U-Expl})~-- for this element we have to verify
an identity. We will come back to it later. Since we know~(\ref{I1}),
(\ref{I2}) and~(\ref{I3}), we may write for all other elements (which
are always governed by expressions with {\it four} terms)

\begin{widetext}
\begin{equation}
\label{I-All-rule1}
2m(m-1)\cdot X+\left[2p+8pm-2m(2m-1)\right]\frac{k+2}{N(N-1)}+\left[4p(p-1)-2p-8pm\right]\frac{k+1}{N(N-1)}-4p(p-1)\frac{k}{N(N-1)}=1,
\end{equation}
\end{widetext}
with $k$ showing the number of the element in the column and starting
from $1$. From equation~(\ref{I-All-rule1}) it follows, that

\begin{equation}
\label{I-All}
X=\frac{k+3}{N(N-1)}.
\end{equation}

Having identified all the elements, all we have to do in the second column
is to check an identity for the first element. This element corresponds to
the operator with $m=0;\ p=N/2$, therefore from expressions
(\ref{gamma-F-fixed-App}) it follows, that the equivalent of
(\ref{I-All-rule1}) for it is

\begin{equation}
\left.\left[4p(p-1)+2p\right]\cdot\frac{1}{N(N-1)}\right|_{p=N/2}=1.
\end{equation}
This is the aforementioned identity for the first element, and it appears
to be true.

This is all we needed to prove for the elements of the second column:
from the expressions~(\ref{I1}),~(\ref{I2}),~(\ref{I3}) and~(\ref{I-All})
it follows, that all elements have the same denominators, namely $N(N-1)$,
and their numerators are equal to $k$, where $k=1$ corresponds to the second
element from the end. All these elements are obtained from the requirement,
that by multiplying the matrix $U_N$, which brings the matrix of
critical dimensions $A_N$ to the Jordan form, with the matrix 
$A_N$ one obtains the matrix $U_N$, but with all
columns shifted by one position to the right. Furthermore, the previous
column is constructed from all the elements, which are equal to~$1$
(see Appendix~\ref{sec:I-Col}).

Moreover, all the formulae~(\ref{I1}),~(\ref{I2}),~(\ref{I3}) and~(\ref{I-All})
can be unified using combinations: since we are dealing with a column
with ${\cal C}=1$, therefore

\begin{equation}
\label{I-All-Comb}
X=\frac{1}{N(N-1)}\cdot {p\choose 1}.
\end{equation}
\subsubsection{The three lower diagonals}

Now we know the form of the elements from the first and the second columns.
This is our starting point for proving the form of all the other elements
in a matrix with finite, but arbitrary number of columns. This will be done
in two steps: first we will prove this for the three lowest diagonals, and
then, in the next section, we will consider all the other elements. The
three lowest diagonals are considered separately since the equations which
determine the elements in these diagonals contain different number of terms.
This situation is similar to the case in the previous section, in which we
considered the first three expressions,~(\ref{I1-rule}),~(\ref{I2-rule}) and~(\ref{I3-rule}),
separately from the general expression~(\ref{I-All-rule1}).

First let us consider the lowest diagonal. The product of each element from
it with the corresponding element of the matrix $A_N$ has to result in the
element with the same position of the previous column of matrix $U_N$.

At this point we know all the elements from the two first columns.
Hence we may start from the last element of the already known column
with ${\cal C}=1$ and then describe the sequence of all other elements
from the lowest diagonal.

The rule, which these elements are governed by, is

\begin{equation}
\label{D1-rule}
X\cdot\gamma_{Np,\,Np'+1}^*=Y.
\end{equation}
Here $X$ and $Y$ denote the elements of the diagonal in question,
but $Y$ is already known element from the column with ${\cal C}_Y=i$
and $X$ is a sought-for element from the column with ${\cal C}_X=i+1$.

According to expression~(\ref{gamma-p+1-App}), the element $\gamma_{Np,\,Np'+1}^*$
is equal to $2m(2m-1)$. Let us start from the elements of the two first
columns, i.e., ${\cal C}_Y=0$ and ${\cal C}_X=1$. In this case
$Y=1$ (see Appendix~\ref{sec:I-Col}) and $2m=N$. Therefore,

\begin{equation}
X= \frac{1}{N(N-1)}.
\end{equation}

The vertical position of each following element in the diagonal is higher
than of the previous, therefore the number $2m$ decreases from $N$ (column
with ${\cal C}=1$) to $2$ (the latest column). From the expression
(\ref{D1-rule}) it then follows, that the sequence of the elements in this
(the lowest) diagonal is

$$\frac{1}{N(N-1)};\quad\frac{1}{N(N-1)(N-2)(N-3)};\quad \dots$$
\begin{equation}
\label{D1-Ans}
\dots\quad \frac{1}{N(N-1)(N-2)(N-3)\cdot .\ .\ . \cdot2\cdot1}.
\end{equation}

For the elements of the second from the bottom diagonal the equation like
(\ref{D1-rule}) takes the form

\begin{widetext}
\begin{equation}
\label{D2-rule}
X\cdot\gamma_{Np,\,Np'+1}^*+ \frac{1}{N(N-1)...(N-2p+2)(N-2p+1)}
\cdot\gamma_{Np,\,Np'}^* = \frac{p}{N(N-1)...(N-2p+4)(N-2p+3)},
\end{equation}
\end{widetext}
where $X$ is the sought-for element and $\gamma_{Np,\,Np'+1}^*$, $\gamma_{Np,\,Np'}^*$
are defined in~(\ref{gamma-p+1-App}) and~(\ref{gamma-p-App}). The numerator
of the RHS follows from the explicit form of those equations. For example,
if ${\cal C}=1$, the sought-for element corresponds to the operator with
$p=2$, and accordingly to~\eqref{I2} the RHS is equal to $2/N(N-1)$. The
solution of this equation, namely expression~\eqref{D2-Ans}, is proportional
to $p+1$ and is the starting point for the next element of the diagonal,
which corresponds to the operator with $p=3$; etc. Note, that the RHS in
the~\eqref{D2-rule} is a {\it known}, but not a sought-for quantity. From
the expression~\eqref{D2-rule} it follows, that

\begin{equation}
\label{D2-Ans}
X=\frac{p+1}{N(N-1)\dots(N-2p+1)},
\end{equation}
in agreement with~(\ref{U-Expl}). Moreover, as we investigate the elements
from the second from the bottom diagonal, the numerator in~\eqref{D2-Ans}
may be written as

\begin{equation}
\label{D2-Ans-Comb}
p+1={p+1\choose p}.
\end{equation}

For the elements of the third from the bottom diagonal the corresponding
expression similar to~(\ref{D1-rule}) and~(\ref{D2-rule}) is

\begin{widetext}
$$
X\cdot\gamma_{Np,\,Np'+1}^*+ \frac{p}{N(N-1)\dots(N-2p+4)(N-2p+3)}
\cdot\gamma_{Np,\,Np'}^*+ \frac{1}{N(N-1)\dots(N-2p+4)(N-2p+3)}
\cdot\gamma_{Np,\,Np'-1}^* =
$$

\begin{equation}
\label{D3-rule}
= \frac{\alpha}{N(N-1)\dots(N-2p+6)(N-2p+5)},
\end{equation}
\end{widetext}
where $X$ is the sought-for element and $\gamma_{Np,\,Np'+1}^*$, $\gamma_{Np,\,Np'}^*$,
$\gamma_{Np,\,Np'-1}^*$ are defined in~(\ref{gamma-p+1-App}),~(\ref{gamma-p-App})
and~(\ref{gamma-p-1-App}). In addition,

\begin{equation}
\label{Alpha-App}
\alpha=3+\sum_{n=3}^{p-1}n=\frac{1}{2}p(p-1).
\end{equation}
From expressions~(\ref{D3-rule}) and~(\ref{Alpha-App}) it follows, that

\begin{equation}
\label{D3-Ans}
X=\frac{\frac{1}{2}p(p+1)}{N(N-1)\dots(N-2p+3)},
\end{equation}
which also may be written using combinations, namely

\begin{equation}
\label{D3-Ans-Comb}
X=\frac{1}{N(N-1)\dots(N-2p+3)}\cdot {p+1\choose p-1}.
\end{equation}

So, we know at this point all the elements from the first two columns and
three lowest diagonals, which satisfy the general requirement that the
product of $A_N\cdot U_N$ be the matrix $U_N$ with all columns, shifted by
one position to the right.

\subsubsection{All other elements}
\label{sec:All-Elements}

Let us introduce some new notations to use only in this subsection.
Let us denote the number of the column as ${\cal C}$, and numeration goes from left to right and starts from ${\cal C}=0$.
Let ${\cal C}_{\cal L}$ be an element from
column ${\cal C}$ with position ${\cal L}$.
The hypotesis is that the numerator of the
element ${\cal C}_{\cal L}$ is the combination ${{\cal L}\choose {\cal C}}$
for all ${\cal C}$, ${\cal L}$. Then, we will use trivial relations
for the combinations, namely

\begin{subequations}
\label{Combin}
\begin{align}
\label{Combin-hor}
{{\cal L}\choose {\cal C}}&={{\cal L}\choose
{\cal C}-1}\cdot\frac{{\cal L}+1-{\cal C}}{{\cal C}},\\
\label{Combin-diag}
{{\cal L}+{\cal C}\choose {\cal C}}&={{\cal L}+{\cal C}-1\choose
{\cal C}-1}\cdot\frac{{\cal L}+{\cal C}}{1+{\cal C}},\\
\label{Combin-vert}
{{\cal L}\choose {\cal C}}&={{\cal L}-1\choose {\cal C}}
\cdot\frac{{\cal L}}{{\cal L}-{\cal C}},
\end{align}
\end{subequations}
where ${\cal C}$ denotes the number of the column and ${\cal L}$ denotes the
number of the row (starts also from ${\cal L}=0$ and with numeration going
from bottom up). Expressions~(\ref{Combin-hor})--(\ref{Combin-vert})
allow us to move in the horizontal, diagonal and vertical directions in the
matrix $U_N$.

The basic equation in the general case is
\begin{widetext}
$$
(N-2p)(N-2p-1)\cdot X+
\frac{2p+4p(N-2p)-(N-2p)(N-2p-1)}{N(N-1)\dots(N-2{\cal C}+1)}
\cdot {\cal C}_{{\cal L}+2}+
$$

\begin{equation}
\label{All-Eq}
+\frac{4p(p-1)-2p-4p(N-2p)}{N(N-1)\dots(N-2{\cal C}+1)}\cdot {\cal C}_{{\cal L}+1}+
\frac{-4p(p-1)}{N(N-1)\dots(N-2{\cal C}+1)}\cdot {\cal C}_{{\cal L}}=\frac{\left({\cal C}-1\right)_{{\cal L}+3}}{N(N-1)\dots(N-2{\cal C}+3)},
\end{equation}
\end{widetext}
where $X$ is the sought-for element. Now we have to verify two hypotheses:

(1) The denominator of the element $X$ is the product
$N(N-1)\dots(N-2{\cal C}+1)$.

(2) The numerator of the element $X$, denoted as ${\cal C}_{{\cal L}+3}$,
is the corresponding combination. Note that we know all the elements of
the three lowest diagonals (see~(\ref{D1-Ans}),~(\ref{D2-Ans}) and
(\ref{D3-Ans})), which are also combinations.

Therefore we want to check, whether the equation~(\ref{All-Eq}) is true if
$X$ satisfies the above written conditions, i.e., if

\begin{equation}
\label{Y-Hyp}
X=\frac{{\cal C}_{{\cal L}+3}}{N(N-1)\dots(N-2{\cal C}+1)}
\end{equation}
and all ${\cal C}$ in~(\ref{All-Eq}) are some combinations.

To verify that let us substitute~(\ref{Y-Hyp}) into~(\ref{All-Eq}).
We find, that

\begin{widetext}
$$
\frac{(N-2p)(N-2p-1)}{(N-2{\cal C}+2)(N-2{\cal C}+1)}\cdot {\cal C}_{{\cal L}+3}+
\frac{2p+4p(N-2p)-(N-2p)(N-2p-1)}{(N-2{\cal C}+2)(N-2{\cal C}+1)}\cdot
{\cal C}_{{\cal L}+2}+
$$

\begin{equation}
\label{All-Eq-Late}
+\frac{4p(p-1)-2p-4p(N-2p)}{(N-2{\cal C}+2)(N-2{\cal C}+1)}\cdot
{\cal C}_{{\cal L}+1}+
\frac{\left[-4p(p-1)\right]}{(N-2{\cal C}+2)(N-2{\cal C}+1)}\cdot
{\cal C}_{{\cal L}}=\left({\cal C}-1\right)_{{\cal L}+3}.
\end{equation}
\end{widetext}
Then, let us express ${\cal C}_{{\cal L}+3}$, ${\cal C}_{{\cal L}+2}$,
${\cal C}_{{\cal L}+1}$
and $\left({\cal C}-1\right)_{{\cal L}+3}$ through ${\cal C}_{{\cal L}}$
using the expressions (\ref{Combin}):
\[
{\cal C}_{{\cal L}+3}=\frac{({\cal L}+2+{\cal C})\,
({\cal L}+1+{\cal C})\,({\cal L}+{\cal C})}
{({\cal L}+2)({\cal L}+1){\cal L}}\cdot {\cal C}_{{\cal L}}; \]
\[
{\cal C}_{{\cal L}+2}=\frac{({\cal L}+1+{\cal C})\,({\cal L}+
{\cal C})}{({\cal L}+1){\cal L}}\cdot
{\cal C}_{{\cal L}}; \]

\[
{\cal C}_{{\cal L}+1}=\frac{({\cal L}+{\cal C})}{{\cal L}}\cdot
{\cal C}_{{\cal L}};\]
\begin{equation}
\left({\cal C}-1\right)_{{\cal L}+3}=\frac{{\cal C}\,({\cal L}+1+{\cal C})\,
({\cal L}+{\cal C})}{({\cal L}+2)({\cal L}+1){\cal L}}\cdot
{\cal C}_{{\cal L}}.
\label{Ck-C}
\end{equation}
Substituting~(\ref{Ck-C}) into~(\ref{All-Eq-Late}) gives us an expression
without the arbitrary number ${\cal C}_{{\cal L}}$, namely

\begin{widetext}
$$
\frac{(N-2p)(N-2p-1)}{(N-2{\cal C}+2)(N-2{\cal C}+1)}\cdot
\frac{({\cal L}+2+{\cal C})\,({\cal L}+1+{\cal C})\,({\cal L}+{\cal C})}{({\cal L}+2)({\cal L}+1){\cal L}}+
$$

$$
+\frac{2p+4p(N-2p)-(N-2p)(N-2p-1)}{(N-2{\cal C}+2)(N-2{\cal C}+1)}\cdot
\frac{({\cal L}+1+{\cal C})\,({\cal L}+{\cal C})}{({\cal L}+1){\cal L}}+
$$

\begin{equation}
\label{All-Eq-Final}
+\frac{4p(p-1)-2p-4p(N-2p)}{(N-2{\cal C}+2)(N-2{\cal C}+1)}\cdot
\frac{({\cal L}+{\cal C})}{{\cal L}}+
\frac{\left[-4p(p-1)\right]}{(N-2{\cal C}+2)(N-2{\cal C}+1)}=
\frac{{\cal C}\,({\cal L}+1+{\cal C})\,({\cal L}+{\cal C})}{({\cal L}+2)({\cal L}+1){\cal L}}.
\end{equation}
\end{widetext}
Moreover, it is obvious that the numbers ${\cal L}$, ${\cal C}$ and $p$ are not
independent: there is a relation between them, namely

\begin{equation}
\label{Relat-Final}
1+{\cal L}+{\cal C}=p.
\end{equation}
Using~(\ref{Relat-Final}) one may check, that the expression
(\ref{All-Eq-Final}) is true, i.e., it is an identity. This means
that our conjecture~(\ref{Y-Hyp}) is true!

\subsection{Summary}

In sections~\ref{sec:I-Col}--\ref{sec:All-Elements} we
proved the hypothesis, that there exists a matrix of special type
that brings our matrix of critical dimensions to Jordan form.
We presented the explicit form of that
matrix $U_N$, for arbitrary dimension $N$ of the set of mixing operators.
As a consequence, the critical dimension matrix~(\ref{Crit-Dim-F})
is degenerate, with known eigenvalues. This leads to violation of scaling
and logarithmic corrections in various correlation functions.




\end{document}